\documentclass[utf8, noinfo]{template/Harvard} 

\usepackage{url}
\usepackage{hyperref}
\usepackage{lineno}
\usepackage{microtype}
\usepackage{subcaption}
\usepackage{siunitx}
\usepackage{verbatim}                   
\usepackage{csquotes} 
\usepackage{tabularx} 
\usepackage{booktabs}
\usepackage{tabu}                       
\usepackage{multirow}                   
\usepackage{balance}
\usepackage{stfloats}
\usepackage[T1]{fontenc}
\usepackage{upgreek}
\usepackage[english]{babel}              
\usepackage{array} 

\usepackage{enumitem}
\usepackage[onehalfspacing]{setspace}

\usepackage[commandnameprefix=ifneeded, final]{changes}
\definechangesauthor[name=R1, color=Green]{R1}
\definechangesauthor[name=R2, color=NavyBlue]{R2}
\definechangesauthor[name=authors, color=BurntOrange]{A}

\usepackage{draftwatermark}
\SetWatermarkLightness{0.85}
\SetWatermarkText{preprint}

\def\keyFont{\fontsize{8}{11}\helveticabold }
\def\firstAuthorLast{Mal {et~al.}} 
\def\Authors{David Mal\,$^{1,*}$, Nina Döllinger\,$^{2}$, Erik Wolf\,$^{1,3}$, Stephan Wenninger\,$^{4}$, Mario Botsch\,$^{4}$, Carolin Wienrich\,$^{2}$, and Marc Erich Latoschik\,$^{1}$}

\def\Address{$^{1}$Human-Computer Interaction Group, University of Würzburg, Würzburg, Germany \\
$^{2}$Psychology of Intelligent Interactive Systems Group, University of Würzburg, Würzburg, Germany \\
$^{3}$Human-Computer Interaction, Universität Hamburg, Hamburg, Germany \\
$^{4}$Computer Graphics Group, TU Dortmund University, Dortmund, Germany}

\def\corrAuthor{David Mal}
\def\corrEmail{david.mal@uni-wuerzburg.de}

\addto\extrasenglish{

}

\begin{document}
\onecolumn
\firstpage{1}

\title[(In)Congruent Realism of Avatars and Virtual Others]{Am I the Odd One? Exploring (In)Congruencies in the Realism of Avatars and Virtual Others in Virtual Reality} 

\author[\firstAuthorLast ]{\Authors} 
\address{} 
\correspondance{} 

\extraAuth{}%

\maketitle
\begin{abstract}
Virtual humans play a pivotal role in social virtual environments, shaping users' VR experiences. The diversity in available options and users' individual preferences can result in a heterogeneous mix of appearances among a group of virtual humans. The resulting variety in higher-order anthropomorphic and realistic cues introduces multiple (in)congruencies, eventually impacting the plausibility of the experience. 
However, related work investigating the effects of being co-located with multiple virtual humans of different appearances remains limited. 
In this work, we consider the impact of (in)congruencies in the realism of a group of virtual humans, including co-located others~(agents) and one’s self-representation~(self-avatar), on users' individual VR experiences. 
In a $2\times3$\, mixed design, participants embodied either (1) a personalized realistic or (2) a customized stylized self-avatar across three consecutive VR exposures in which they were accompanied by a group of virtual others being either (1)~all realistic, (2)~all stylized, or (3)~mixed between stylized and realistic. 
Our results indicate groups of virtual others of higher realism, i.e., potentially more congruent with participants’ real-world experiences and expectations, were considered more human-like, increasing the feeling of co-presence and the impression of interaction possibilities. (In)congruencies concerning the homogeneity of the group did not cause considerable effects. Furthermore, our results indicate that a self-avatar’s congruence with the participant’s real-world experiences concerning their own physical body yielded notable benefits for virtual body ownership and self-identification for realistic personalized avatars. 
Notably, the incongruence between a stylized self-avatar and a group of realistic virtual others resulted in diminished ratings of self-location and self-identification. This suggests that higher-order (in)congruent visual cues that are not within the ego-central referential frame of one’s (virtual) body, can have an (adverse) effect on the relationship between one's self and body.
We conclude on the implications of our findings and discuss our results within current theories of VR experiences, considering (in)congruent visual cues and their impact on the perception of virtual others, self-representation, and spatial presence. \looseness=-1

\tiny
 \keyFont{ \section{Keywords:} Virtual Human, Virtual Others, Avatar, Agent, Sense of Embodiment, Co-Presence, Plausibility, Presence}

\end{abstract}

\section{Introduction}
Social virtual environments (SVEs) have gained significant attention for their remarkable ability to foster pro-social interactions and push the boundaries of traditional digital collaboration platforms~\citep{mcveigh2019, schulzWelcomeMetaverseComprehensive2021}. Within this realm, virtual humans can play a pivotal role, facilitating a multitude of mixed, augmented, and virtual reality~(MR,~AR,~VR:~XR~for~short) experiences. As so-called avatars, they can be directly controlled and embodied by users \citep{slaterInducingIllusoryOwnership2009, bailensonAvatars2004}, enabling a bodily experience and sharing social signals in SVEs \citep{benteAvatarMediatedNetworkingIncreasing2008, kolesnichenkoUnderstandingEmergingDesign2019, freemanBodyAvatarMe2021}. Yet, the versatility of virtual humans within SVEs transcends self-representation, as they may function as computer-controlled agents \citep{bailensonAvatars2004}, or seamlessly blend into embodied ambient crowds \citep{latoschikNotAloneHere2019}. 
Numerous commercially available SVEs like VRChat, RecRoom, and Mozilla Hubs (as outlined in~\cite{liusteedSocialVirtualReality2021}) provide a broad spectrum of virtual humans varying in their appearances or even allow users to upload their individual avatars~\citep{kolesnichenkoUnderstandingEmergingDesign2019, hepperleAspectsVisualAvatar2022}. Therefore, the styles of virtual humans and their degree of realism and individualization can differ significantly. Realistic virtual humans may resemble life- and human-like visual features \citep{latoschikEffectAvatarRealism2017}, while stylized or abstract virtual humans yield a rather simplified or iconic style~\citep{zellStylizeNotStylize2015, lugrinAnthropomorphismIllusionVirtual2015}. Individualization, on the other hand, tailors a virtual human to a specific individual, fostering the digital representation to become congruent with the user's physical appearance.
Conversely, a more generalized appearance enables playful and creative avatar selection or the maintenance of a desired distance from one's physical body. Users' avatar choices can also be influenced by their access to avatar-creation technologies. While some have the ability to craft personalized, life-like avatars through 3D reconstruction from multi-view scanning~\citep{achenbachFastGenerationRealistic2017} or by using smartphone cameras~\citep{wenningerRealisticVirtualHumans2020}, others may settle for customizing virtual humans or selecting generic ones provided by the application \citep{hepperleAspectsVisualAvatar2022}.
This diversity of options and individual preferences can result in a heterogeneous mix of various appearances among virtual humans in SVEs.\looseness=-1

A substantial body of research has placed emphasis on understanding how the visualization of avatars and agents impacts users' virtual experiences and their evaluation of co-located virtual others \citep{weidnerSystematicReviewVisualization2023, nowakAvatarsComputermediatedCommunication2018}. However, related work investigating the effects of being co-located with multiple virtual humans of different styles seems limited. \cite{latoschikNotAloneHere2019} suggested that employing mixed appearances for multiple virtual others in an ambient crowd could enhance participants' interest in the virtual environment. However, they might also increase feelings of eeriness~\citep{latoschikNotAloneHere2019} and introduce incongruencies within a group of virtual humans. 
In this context, the concepts of congruence and plausibility have emerged as fundamental when exploring VR experiences~\citep{slaterSeparateRealityUpdate2022, latoschik2022plausibility, skarbezImmersionCoherenceResearch2020}. This includes considering virtual humans and their (in)congruent appearance and behavior within a particular virtual environment as an essential feature contributing to a user's plausible VR experience~\citep{skarbezPsychophysicalExperimentRegarding2017, mal2022virtual, wolfPlausibilityPerceptionPersonalized2022}.
Related work has investigated typical cues potentially influencing an (in)congruent appearance and behavior of virtual humans, including various factors of virtual human visualization~\citep{weidnerSystematicReviewVisualization2023} such as realism \citep{latoschikEffectAvatarRealism2017, zibrekPhotorealismImportantPerception2019} or personalization \citep{waltemateImpactAvatarPersonalization2018, fribourgAvatarSenseEmbodiment2020}. In turn, these cues have been shown to potentially influence various qualia reflecting the VR experience, such as spatial presence (the feeling of really being in a VE)~\citep{slaterPlaceIllusionPlausibility2009}, the sense of embodiment (the feeling of being inside, having, and controlling an avatar in a VE)~\citep{slaterInducingIllusoryOwnership2009}, as well as, co-presence (the subjective experience of being in the company of virtual others) \citep{Schroeder2002}. Yet, to our knowledge, there has not been a further investigation into the effects of being co-located with multiple virtual humans of different styles in an avatar-mediated VR setting, considering both the realism of co-located virtual humans,  the realism of the avatar, and their (in)congruencies. We conclude with the following research question:\looseness=-1 

\begin{enumerate}[leftmargin=*,labelindent=16pt,labelsep=8pt,rightmargin=32pt, start=1,label={\textbf{RQ:}\,}]
    \item How do the avatar's realism, co-located virtual humans' realism, and their (in)congruencies affect the perception of virtual others, the self-presentation, and the overall VR experience? \looseness=-1
\end{enumerate}

We investigate the stated research question by focusing on the intricate dynamics emerging from (in)congruent styles of a group of virtual humans, including multiple co-located others (agents), and one's digital self-representation (avatar). Therefore, we conducted a user study in which 48 participants each embodied an individualized avatar with varying degrees of realism, i.e., either being realistic or stylized, while consecutively being accompanied by three groups of virtual others also varying in their realism~(i.e.,~all realistic, all stylized, or mixed). Realistic avatars were personalized by scanning participants with a custom-made photogrammetry rig and applying a 3D-reconstruction photogrammetry pipeline. Stylized avatars were customized by the participants using a lightweight graphical user interface.
Accompanied by virtual others, each participant engaged in three VR exposures, utilizing an odd-one-out logic task paradigm. Embodiment was implemented with a state-of-the-art markerless tracking system. 
We evaluate the impact of the realism of a group of virtual others, the self-avatar realism, and the resulting (in)congruencies on users' VR experiences and discuss the results in the context of current theories and models, providing valuable insights for designing and developing future social virtual environments utilizing virtual humans.\looseness=-1
\section{Related Work}
\label{sec:related-work}

\subsection{Virtual Humans}

Drawing from diverse specifications in previous research~\citep{nowakAvatarsComputermediatedCommunication2018, doernerVirtualAugmentedReality2022, burden2019virtual, kyrlitsiasSocialInteractionAgents2022}, we adopt a conceptual and operational definition of virtual humans as user (avatar) or system (agent) controlled digital representations of human beings in a virtual environment. The characteristics of virtual humans are determined by various technical components 
typically related to their general form (e.g., shape and resolution of mesh, dimensions, and proportions), surface (e.g., topology, shading, and texture), motion (e.g., facial and body animation), and sound (e.g., voice or heartbeat)
~\citep{zellPerceptionVirtualCharacters2019, lugrinAnthropomorphismIllusionVirtual2015, burden2019virtual}, collectively contributing to the life- and human-like qualities of the virtual human's behavior and appearance. \looseness=-1

\subsubsection{Realistic and Anthropomorphic Cues in Virtual Humans}
Deriving from \cite{nowakAvatarsComputermediatedCommunication2018}, we refer to the realism of virtual humans as the perception that they could realistically or possibly exist in a non-mediated context (life-like) and anthropomorphism as the perception or assignment of human traits or qualities to these entities (human-like). We argue that the realism of virtual humans in their literal and conceptual meaning is fundamentally rooted in anthropomorphism, as the life-likeness of a human representation depends on its contingent to incorporate human-like features. Therefore, the realism of virtual humans, as an overarching term, may be characterized by an array of realistic and anthropomorphic cues. 
Further specifying, \cite{lugrinAnthropomorphismIllusionVirtual2015} discerned two categories of anthropomorphic cues in the appearance of virtual humans: anatomy, which details the structure, number, and interconnections of body parts, and composition, which describes the technical properties of specific body parts. 
A classification of cues 
can be the base for comparing and categorizing virtual humans of various artistic styles. In a recent review, \cite{weidnerSystematicReviewVisualization2023} classified realistic visualizations of avatars and agents as detailed models of humans based on real persons and a stylized visualization to maintain human proportions with detailed body parts, yet lacking human-like textures and not necessarily adhering to human morphology. Stylization can also be classified along iconic and non-iconic scales, and the stylization of individual cues of virtual humans may also be considered independently~\citep{zellPerceptionVirtualCharacters2019}, e.g., across shape~\citep{zellStylizeNotStylize2015}, rendering styles~\citep{mcdonnellRenderMeReal2012, zibrekDonStandClose2017, wisessingPerceptionShading2016, zellStylizeNotStylize2015, volanteEffectsVirtualHuman2016}
, or composition~\citep{lugrinAnthropomorphismIllusionVirtual2015}. Overall, \cite{weidnerSystematicReviewVisualization2023} indicates realism in avatars and agents to potentially benefit a multitude of qualia related to the VR experience. \looseness=-1

\looseness=-1

\subsubsection{Individualization and Truthfulness of Virtual Humans}
Virtual humans can differ not only in style but also in whether and how they are individualized, i.e., tailored to a specific individual, and the resulting truthfulness, i.e., the degree of similarity between the user's appearance and the virtual human~\citep{gorisseRobotVirtualDoppelganger2019}. Thereby, we refer to personalization as the process of creating a virtual human based on user (photogrammetric) data \citep{waltemateImpactAvatarPersonalization2018} and to customization as the process through which a user actively alters the visual properties of a virtual human \citep{ducheneautBodyMindStudy2009}. The individualization of virtual humans as self-avatars can facilitate the acceptance of the virtual body as ones own and also enhance the identification with the self-avatar, which applies for personalization~\citep{waltemateImpactAvatarPersonalization2018, gorisseRobotVirtualDoppelganger2019, fiedler2023selfidentification, salageanMeetingYourVirtual2023} as well as customization \citep{leeHowAvatarIdentification2023, dollingerIfItNot2023}.\looseness=-1

\subsection{Exploring (In)Congruencies in a Group of Virtual Humans}

Coherence and plausibility have gained significant attention as fundamental concepts in describing and classifying mixed, augmented, and virtual reality experiences \citep{slaterSeparateRealityUpdate2022,latoschik2022plausibility, skarbezImmersionCoherenceResearch2020}. 
As introduced by \cite{slaterPlaceIllusionPlausibility2009}, plausibility describes the illusion that what is happening is really happening. Coherence, on the other hand, refers to the extent to which a virtual scenario behaves reasonably and predictably, thus creating the illusion of plausibility~\citep{skarbezSurveyPresenceRelated2017}. 
\cite{latoschik2022plausibility} introduced an alternative theoretical Congruence and Plausibility~(CaP) model. It proposes plausibility and congruence to become central conditions in describing XR experiences and effects. According to the CaP model, congruence, as the ontological specification of coherence, describes the objective match between processed and expected cues on the sensory, perceptual, and cognitive layers, creating a state of plausibility that influences various qualia and constructs of XR. With reference to the CaP model, virtual humans, whether as avatars or agents, and the congruence of their cues can also contribute to a user's plausible XR experience. In this regard, \cite{mal2022virtual} framed the subjective feeling of how reasonable and believable a virtual human appears to a user as \textit{virtual human plausibility~(VHP)}. \added[id=R1]{VHP would, therefore, arise from the congruence of habitual sensory, proximal perceptual, or higher-order cognitive cues of the virtual human in the VE and eventually impact various qualia shaping the XR experience.}
In this work, we explore (in)congruent appearances in a group of virtual humans, including co-located others (agents) and one's self-representation~(self-avatar). Therefore, following the CaP model, we will subsequently identify manipulated cues, delineate their (in)congruencies, and pinpoint relevant qualia potentially influenced by the specific manipulation as summarized in \autoref{table:virtual_humans}. \looseness=-1

\begin{table}[b]
\small
\centering
\caption{ Overview specifying our manipulation and related cues for virtual others and the self-avatar in a group of virtual humans, further naming expected (in)congruencies, and defining this works' qualia space. \looseness=-1}
\label{table:virtual_humans}

\begin{tabularx}{\textwidth}{l >{\centering\arraybackslash\hsize=.9\hsize}X>{\centering\arraybackslash\hsize=0.9\hsize}X>{\centering\arraybackslash\hsize=0.9\hsize}X>{\centering\arraybackslash\hsize=1.3\hsize}X}
\toprule
      & \multicolumn{2}{c}{Manipulation} & \multirow{2}{*}{(In)Congruencies} & \multirow{2}{*}{Qualia} \\
  
     \cmidrule(l{1pt}r{7pt}){2-3}
   &  Specification & Related Cues & \\
   
\midrule
 \textbf{Virtual Others} & Realistic, Mixed, Stylized  & \parbox[t]{0.9\linewidth}{\centering Realism,\\Anthropomorphism}   & Lifelikeness, Homogeneity, Self-Avatar & \parbox[t]{1.3\linewidth}{\centering Virtual Human Plausibility,\\Co-Presence,\\Affective Appraisal,\\Spatial Presence} \vspace{0.5em}\\ 
 
  \textbf{Self-Avatar} & Realistic, Stylized & \parbox[t]{0.9\linewidth}{\centering Realism,\\Anthropomorphism,\\Truthfulness}  & \parbox[t]{0.9\linewidth}{\centering Physical Body, \\ Virtual Others}  & \parbox[t]{1.3\linewidth}{\centering Sense of Embodiment,\\Self-Identification,\\Spatial Presence}  \\
 
\bottomrule
\end{tabularx}
\end{table}

\subsubsection{Manipulation Space}
We classify our manipulation of appearance on a common scale between realism and stylization \citep{zellPerceptionVirtualCharacters2019}. Therefore, we chose two types of virtual humans distinct in cues related to their form and texture: \looseness=-1

\begin{enumerate}[leftmargin=*,labelindent=16pt,labelsep=8pt,rightmargin=32pt, start=1,label={(\arabic*)}]
    \item Realistic virtual humans created by a 3D-reconstruction photogrammetry process striving for life and human-like appearance.
    \item Stylized virtual humans of lower realism and anthropomorphism, based on a cartoon-style 3D model with human anatomy but simplified composition.
\end{enumerate}

As for the group of others, we manipulated the group's configuration to be either realistic or stylized, i.e., a homogeneous group of virtual humans, or mixed, i.e., a group evenly distributed between stylized and realistic virtual humans. Furthermore, self-avatars were either realistic, based on scan-based personalization, or stylized, based on customization \replaced[id=R1]{. We}{ constrained by available customization options. Here, we} assume\replaced[id=R1]{ realistic avatars to be of a higher level of truthfulness as they objectively resemble more visual features similar to the user's real appearance than the stylized ones. Within the technical system's boundaries (see \autoref{subsubsec:realistic}), the scan-based avatars are created using users' photogrammetric data, depicting the user in form and texture. On the other hand, customized stylized avatars are constrained by available customization options and a simplified composition.}{the personalization process will result in a higher degree of truthfulness compared to the customization process.}\looseness =-1

\subsubsection{Condition Space}
\label{subsubsec:condition}
Drawing from the CaP model, we hypothesize the named manipulation of realism across a group of virtual humans to result in multiple (in)congruencies in higher-order visual cues \added[id=R2]{on a cognitive layer}. 

\begin{enumerate}[leftmargin=*,labelindent=16pt,labelsep=8pt,rightmargin=32pt, start=1,label={(\arabic*)}]
    \item (In)congruencies within the group of virtual others depending on its style configuration. 
    Overall, a homogeneous group configuration appears more congruent than a mixed one, an (in)congruence that can be accessed by directly comparing others within the VE. Furthermore, we assume the realism of each group member to shape the entire group's congruence with the participant's real-world experiences and expectations. With the highest realism/congruence for realistic others, less for mixed, and lowest for stylized others.  \looseness=-1
    \item  The self-avatar's (in)congruence with the participant's real-world experiences concerning their own physical body. We expect a personalized realistic avatar to be of higher realism and truthfulness and, therefore, more congruent with participants' experiences and expectations towards their physical body than a customized stylized avatar. \looseness=-1
    \item The group of virtual others' (in)congruence with the self-avatar, and vice versa. We suggest the self-avatar to be congruent with a homogeneous group of the same style, less congruent with mixed groups, and incongruent with the homogeneous group of the opposite style.\looseness=-1
\end{enumerate}

\subsubsection{Qualia Space}
While the CaP model defines the overall frame for manipulating cues and their congruencies, it does not provide a ranking of (in)congruencies or specify changes in the qualia space. Therefore, in the following, we identify relevant qualia and constructs describing the VR experience and, thereby, review related research indicating how the named (in)congruencies may influence the perception of co-located virtual others, the perception of self-representation, and the VR experience.

\subsection{(In)Congruecies and the Perception of Virtual Others}
Prior research has examined how the congruence in various sensory impressions of virtual others affects the XR experience. The work of \cite{skarbezPsychophysicalExperimentRegarding2017} investigated a virtual agent's behavior coherence and its relative importance as a contributing factor for the overall plausibility of a VR experience. Further research focused on the congruence of spatial and behavioral cues of agents~\citep{kimEffectsVirtualHuman2017}, gaze behavior, and auditory features of virtual groups \citep{bergstromPlausibilityStringQuartet2017}, facial animation methods~\citep{kullmannEvaluationOtherAvatarFacial2023}, different virtual body animation features for avatars and agents~\citep{debarbaPlausibilityVirtualBody2021}, and renderings of single virtual humans and their (in)congruence with the device-related presentation of the respective environment~\citep{wolfPlausibilityPerceptionPersonalized2022}.  
However, we are unaware of related studies investigating the congruence of styles within a group of virtual humans. Following \cite{mal2022virtual}'s conceptualization of VHP, \deleted[id=R1]{we hypothesize} (in)congruencies in the realism of virtual others and their (in)congruence with the \added[id=R1]{self-}avatar \replaced[id=R1]{might}{to} impact the perceived plausibility of virtual others'\added[id=R1]{. We hypothesize more congruent conditions to result in a higher attribution of plausibility towards others} and deduce the following hypotheses.\looseness=-1

\begin{enumerate}[leftmargin=*,labelindent=16pt,labelsep=8pt,rightmargin=32pt, start=1,label={H\arabic*:\,}]
    \item \added[id=R2]{The manipulation of virtual others' realism and the self-avatar's realism will lead to significantly higher scores in VHP for configurations of higher congruence.} 
\end{enumerate}

\subsubsection{Co-Presence and the Impression of Interaction Possibilities}
Co-presence describes the subjective experience of being in the company of others in a virtual environment, or in short, a sense of \enquote{being there together}~\citep{Schroeder2002}. We refer to it as a quale, denoting the sensation of being together in a (virtual) place~\citep{skarbezSurveyPresenceRelated2017}. As noted by \cite{kyrlitsiasSocialInteractionAgents2022}, co-presence (referred to as social presence by the authors) can enhance the realism and effectiveness of interactions between users and virtual humans. 
\added[id=R1]{One decisive feature influencing co-presence in virtual environments is the visual representation of others.}
While the congruence between realism in appearance and behavior has been named of great importance for co-presence ~\citep{garauImpactAvatarRealism2003, bailensonIndependentInteractiveEffects2005}, \replaced[id=R1]{previous}{early} work 
provided mixed results about the significance of a realistic or anthropomorphic appearance of \added[id=R1]{avatars and} agents\deleted[id=R1]{ \mbox{\citep{bailensonEquilibriumTheoryRevisited2001, NowakAnthropomorphism}.}}\added[id=R1]{ \mbox{\citep{ohSystematicReviewSocial2018}}. For example, }
in the work of \cite{latoschikEffectAvatarRealism2017}, the realism of avatars (abstract / realistic) in dyadic VR scenarios did not affect co-presence, though the authors reported slightly higher eeriness ratings for the realistic avatars potentially indicating an Uncanny Valley effect. In contrast, \cite{zibrekDonStandClose2017} found an impact of the rendering styles of animated agents on co-presence in two VR experiments. Interestingly, realistic virtual humans were preferred over stylized representations, while the rendering styles were rated comparable in unappealing and eeriness.
\added[id=R2]{Also, \mbox{\cite{volanteEffectsVirtualHuman2016}} indicated interactions with virtual patients of realistic rendering to increase emotional bonding and social presence compared to stylized ones.} 
For multiple virtual others, \cite{latoschikNotAloneHere2019} indicated the impression of interaction possibilities~(IIP) and co-presence to be consistently higher for a mixed ambient crowd of abstract and realistic virtual participants' along with increased feelings of eeriness. The authors reason that potential incongruencies with participants' expectations may have caused them to focus more intensely on the surrounding agents when they were of mixed appearance. 
Lastly, the realism in shape and texture of the self-representation and its congruence with the appearance of an agent was not found to impact co-presence \added[id=R1]{in the work of} \cite{latoschikEffectAvatarRealism2017}. \added[id=R1]{We did not find further indication for the (in)congruence in realism between the self-avatar and others to impact co-presence and IIP.}
\replaced[id=R1]{In conclusion, we acknowledge that related work on co-presence, specifically investigating the effects of being co-located with multiple virtual humans of different styles in immersive VR, is limited and rather inconclusive. Therefore, we concur with the systematic review by \cite{ohSystematicReviewSocial2018}, which affirms mixed results regarding the impact of others' realism on co-presence. We propose an exploratory evaluation of how our manipulation affects co-presence and IIP. }
{In conclusion, related work identified the realism in virtual others to impact co-presence with mixed results, while an incongruent style configuration for ambient crowds indicated higher co-presence and IIP. Furthermore, we did not find indications for the (in)congruence in realism between the self-avatar and others to impact co-presence and IIP.} 
\looseness=-1 

\subsubsection{Affective Appraisal and the Uncanny Valley Effect}
As indicated, incongruencies in realistic and anthropomorphic cues of virtual humans might lead to increased feelings of eeriness or unappealing towards virtual humans. In this regard, the Uncanny Valley effect delineates a phenomenon describing a transition from an initial affinity to a feeling of eeriness as the appearance of a robot or any anthropomorphic character approaches a convincingly human-like representation but falls short of achieving it \citep{moriUncannyValleyField2012}. There have been indications of an uncanny valley effect in VR for anthropomorphic virtual characters of different visual realism or humanness~ \citep{latoschikEffectAvatarRealism2017, hepperleAspectsVisualAvatar2022}. 
To control for the potential impact of an Uncanny Valley effect, or a general sense of discomfort towards virtual others, we propose an exploratory evaluation of how their eeriness is perceived. Additionally, we aim to assess our manipulation in terms of perceived humanness.\looseness=-1

\subsection{(In)Congruencies and the Perception of Self-Representation}

\subsubsection{Sense of Embodiment \added[id=R2]{and Self-Identification}}
An essential quale describing the experience of having an avatar is the sense of embodiment~(SoE). It emerges when a virtual body's properties are processed as if they were the properties of one's own physical body \citep{kilteniSenseEmbodimentVirtual2012}. SoE has been named an essential concept for other effects like the Proteus effect to emerge \citep{malImpactAvatarEnvironment2023, ratan2019proteus}. It is considered to consist of three sub-dimensions describing the senses of having (body ownership), controlling~(agency), and being inside~(self-location) a virtual body~\citep{longoWhatEmbodimentPsychometric2008, slaterFirstPersonExperience2010, kilteniSenseEmbodimentVirtual2012}. \cite{roth2020construction} further named a perceived change in the body schema as an essential component of the SoE, a factor that is rather significant for studies using altered body appearances, according to the authors. \added[id=R2]{
The named sub-components can be assessed with well-established questionnaires \mbox{\citep{roth2020construction, peckAvatarEmbodimentStandardized2021}.}}
In general, the SoE is understood to stem from integrating both bottom-up and top-down influences \citep{kilteniSenseEmbodimentVirtual2012, maselliBuildingBlocksFull2013}. Bottom-up information arises from the congruence of visual, tactile, and proprioceptive cues, predominantly determined by the capabilities of the technical system \citep{kilteniSenseEmbodimentVirtual2012, slaterSeparateRealityUpdate2022}. On the other hand, top-down information is derived from the cognitive processing of the avatar's visual cues and their congruencies, encompassing factors like self-similarity \citep{fiedler2023selfidentification, waltemateImpactAvatarPersonalization2018, salageanMeetingYourVirtual2023} and realism \citep{latoschikEffectAvatarRealism2017, salageanMeetingYourVirtual2023}.

Regarding the visualization of avatars, a recent review by \cite{weidnerSystematicReviewVisualization2023} indicated that virtual body ownership (VBO) indeed benefits from both personalized and realistic appearances. This suggests that congruence between users' real-world experiences, particularly with their own physical bodies, and the self-avatar can enhance the sense of owning and accepting a virtual body. For our manipulation, we expect (personalized) realistic self-avatars to be more congruent with the participant's physical appearance, potentially leading to higher VBO values.
Conversely, a current meta-analysis by \cite{mottelsonSystematicReviewMetaanalysis2023} pointed out that the appearance of the self-avatar seems to be of less importance for the perceived agency over the virtual body. We concur with this conclusion, provided that the underlying technical system ensures comparable contingencies for visuomotor congruence across the self-avatar types, a condition we also assume to be met by our embodiment system. 
For self-location, we follow \cite{kilteniSenseEmbodimentVirtual2012}, naming it to be primarily influenced by the visuospatial perspective, another factor we kept constant for all conditions in our system configuration. 

\added[id=R2]{Another important component concerning the perception of self-avatars is the process of identifying with the digital representation, namely self-identification~\mbox{\citep{gonzlezfranco2020avatarselfidentification}}. Interestingly, the body of research focused on self-identification remains limited \mbox{\citep{fiedler2023selfidentification}}, despite its recognized significance as a pivotal factor driving relevant XR effects, like the Proteus effect \mbox{\citep{ratan2019proteus}}.  Following \mbox{\cite{wolf2022observationdistance}}, self-identification encompasses two key components: self-similarity as the perceived visual congruence between the individual and the virtual human, and self-attribution as the attribution of personal characteristics to the virtual human for both external body features or internal character traits. \cite{salageanMeetingYourVirtual2023} recently investigated self-identification towards virtual humans with a full-body embodiment system, reporting higher realism and personalization to be beneficial for self-identification. We attribute these findings to higher-order visual cues determining the congruence between the participant's physical body and the self-avatar.} 

We conclude with the following hypotheses \added[id=R2]{for the sense of embodiement and self-identification}.

\begin{enumerate}[leftmargin=*,labelindent=16pt,labelsep=8pt,rightmargin=32pt, start=1,label={H2.\arabic*:\,}]
    \item The manipulation of the self-avatar's congruence with the participant will lead to significantly higher scores in VBO \added[id=R2]{and self-identification} for realistic self-avatars.   
    \item The manipulation of the self-avatar's congruence with the participant will not significantly affect agency, change, and self-location. 
\end{enumerate}	 
 
\deleted[id=R2]{Another important component concerning the perception of self-avatars is the process of identifying with the digital representation, namely self-identification~\mbox{\citep{gonzlezfranco2020avatarselfidentification}}. Interestingly, the body of research focused on self-identification remains limited \mbox{\citep{fiedler2023selfidentification}}, despite its recognized significance as a pivotal factor driving relevant XR effects, like the Proteus effect \mbox{\citep{ratan2019proteus}}. 
Following \mbox{\cite{wolf2022observationdistance}}, self-identification encompasses two key components: self-similarity as the perceived visual congruence between the individual and the virtual human, and self-attribution as the attribution of personal characteristics to the virtual human for both external body features or internal character traits. 
\mbox{\cite{salageanMeetingYourVirtual2023}} recently investigated self-identification towards virtual humans with a full-body embodiment system, reporting higher realism and personalization to be beneficial for self-identification. We attribute these findings to 
higher-order visual cues determining the congruence between the participant's physical body and the self-avatar. We conclude with the following hypothesis.}

\subsubsection{The \added[id=R2]{Perception of Self-Representation} \deleted[id=R2]{Congruence Between the Self-Avatar} and Virtual Others}
Interestingly, \cite{latoschikEffectAvatarRealism2017} found a marginal effect of an agent's realism (realistic vs. abstract) on embodied users' change in self-perceived body schema \added[id=R2]{(change)}. This suggests that the appearance of virtual others can extend beyond affecting general user experience or the evaluation of the virtual others. It may also influence self-related concepts within the ego-central referential frame of one's (virtual) body, such as the SoE and self-identification toward an avatar. It is worth noting that insight in exploring these concepts dependent on the visualization of virtual others remains limited \citep{malDCImpactSocial2020}, especially in multi-agent embodied virtual environments. We, therefore, \deleted[id=A]{further} propose an exploratory evaluation of the impact of the realism of virtual others and the congruence between the self-avatar and virtual others on the perception of the self-avatar.\looseness=-1 

\subsection{Spatial Presence}

Spatial presence is an essential concept in evaluating a user's individual VR experience and can be considered an elementary foundation for other VR potentials to become effective \citep{wienrichBehavioralFrameworkImmersive2021}. \replaced[id=R1]{The process model of the formation of spatial presence experiences names self-location as the core dimension of spatial presence, i.e., ``the sensation of being physically situated within the spatial environment portrayed by the medium''~\mbox{\citep[~p.~497]{wirthProcessModelFormation2007}}. However, to avoid ambiguities, we adhere to the definition of \mbox{\cite{kilteniSenseEmbodimentVirtual2012}} and name the term self-location to refer to one’s spatial experience of being inside a (virtual) body rather than being inside a world, whereas we apply the term spatial presence to the (psychological) sense of ``being there'' in a virtual environment~\mbox{\citep{slaterFrameworkImmersiveVirtual1997}}. Spatial presence}{
It's the (psychological) sense of being in a virtual environment~\mbox{\citep{slaterFrameworkImmersiveVirtual1997}}and} has been named to be predominantly bottom-up driven by the objective concept of immersion, which describes the capabilities of the system providing the boundaries within which spatial presence occurs \citep{slaterSeparateRealityUpdate2022}. 
\cite{latoschik2022plausibility}, on the other hand, assumes spatial presence to be affected by the \replaced[id=A]{congruence}{ coherence} and plausibility of spatial cues on the sensation, perception, and cognition levels, not limiting influencing factors to the system's capabilities.
Concerning the avatar's congruence with the physical appearance of the participant, previous work has shown that spatial presence can benefit from personalization~\citep{waltemateImpactAvatarPersonalization2018}. 
However, the recent work of \cite{salageanMeetingYourVirtual2023} did not reveal a significant effect of avatar personalization (high/low), realism (high/low), and their interaction on presence. As these results contradicted the authors' hypotheses, as well as the results of previous work \citep{waltemateImpactAvatarPersonalization2018}, the authors attribute their findings to a constant level of agency and appearance of the virtual environment across all conditions. Also, \cite{latoschikEffectAvatarRealism2017} found no impact of virtual human realism on spatial presence, applying to avatars and agents. Further related work indicates no effect of agent visualization on spatial presence \citep{rzayevEffectPresenceAppearance2019, butzInfluenceVisualAppearance2022}.\looseness=-1

In our experiment, we maintained a constant degree of immersion (bottom-up) across conditions. Furthermore, as argued by \cite{salageanMeetingYourVirtual2023}, we expect a consistent level of agency and a stable appearance of the virtual environment. Additionally, we anticipate no significant manipulation of higher-order spatial cues through the manipulation of virtual humans' realism and their (in)congruencies. Finally, we are not aware of any work indicating that the realism of virtual others significantly impacts spatial presence. Consequently, we propose the following null hypothesis.

\begin{enumerate}[leftmargin=*,labelindent=16pt,labelsep=8pt,rightmargin=32pt, start=1,label={H3:\,}]
    \item The manipulation of the \added[id=R1]{self-}avatar's realism, the virtual others' realism, and their congruence will not significantly affect spatial presence.
\end{enumerate}	 


We conducted a controlled user study to investigate the stated hypotheses and proposed exploratory evaluations. Therefore, we systematically manipulated the realism of a group of virtual others and the self-avatars and evaluated participants' ratings on their VR experience based on their perception of the virtual others, self-representation, and spatial presence. 

\section{Materials and Methods}
\label{sec:methods}

\subsection{Participants}
\label{subsec:participants}
A total of 53 undergraduate students from the University of Würzburg were recruited through a participant management system and received course credit for participation according to the study duration. We excluded data from one participant reporting uncorrected visual impairment and another reporting color blindness. Three data sets were further excluded due to technical issues. For the resulting 48 valid data sets, ages ranged from 18 to 27 years ($M=21.73$, $SD=2.17$), with 24 participants identifying as female and 24 as male. Participants were evenly distributed between conditions, leading to gender-balanced groups of equal sample sizes. All participants were German native speakers. One participant was new to VR, 42 participants reported having less than 10 hours of VR experience, and five participants reported having over 10 hours of VR experience. The study was conducted according to the Declaration of Helsinki and approved by the ethics committee of the Human-Computer-Media Institute of the University of Würzburg.

\begin{figure*}[tb]
  \centering
 \includegraphics[width=\linewidth]{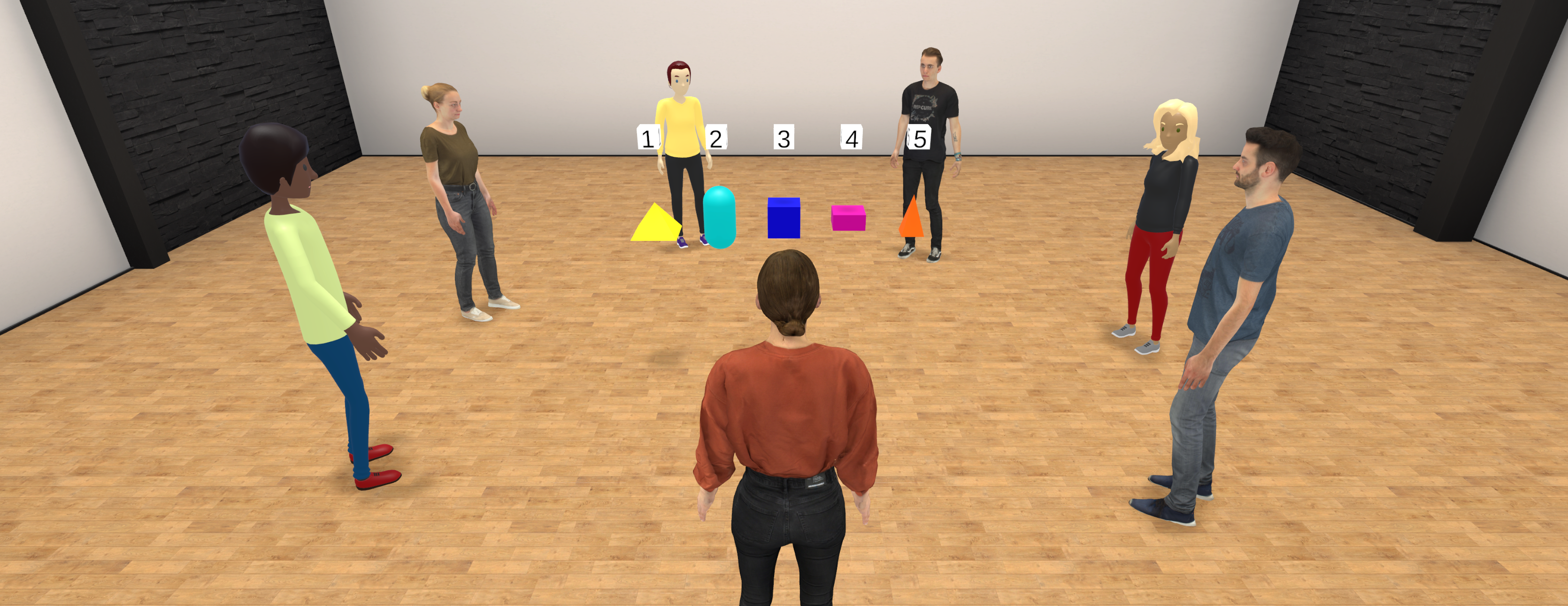}
  \caption{A participant embodying a realistic self-avatar (front) while solving a logic task in a group of virtual others (back). The group features a configuration of mixed styles consisting of three realistic virtual humans and three stylized virtual humans.}
  \label{fig:teaser}
\end{figure*}

\subsection{Design} 
\label{subsec:design}
We employed a $2\times3$ mixed design with the independent variables \textit{self-avatar} (realistic and stylized) as the between-subject factor and \textit{virtual others} (realistic, stylized, and mixed) as the within-subject factor. Therefore, participants either embodied a personalized realistic avatar or a customized stylized avatar in three consecutive VR exposures. Each time, they solved a set of logical tasks in VR accompanied by a group of virtual others varying in their realism, i.e.,~all realistic, all stylized, or evenly distributed between stylized and realistic\replaced[id=R1]{. The within-subject manipulation was presented}{,} in counterbalanced order. Dependent variables assessed the perception of the virtual others after each VR exposure regarding plausibility, affective appraisal, and perceived co-presence and interaction possibilities. We evaluated the individuals' VR experience by assessing spatial presence. Further, we considered the perception of self-presentation in terms of the sense of embodiment and self-identification with the self-avatar after each exposure and once overall after the last exposure.

\subsection{Measures}
\label{subsec:measures}
 
\subsubsection{Plausibility of Virtual Others}
We assessed the virtual others' plausibility with the \emph{Virtual Human Plausibility Questionnaire}~(VHPQ) \citep{mal2022virtual}. It consists of two dimensions: (1) The virtual human's appearance and behavior plausibility~(ABP) and (2)~the virtual humans's match to the VE~(MVE). The 11 items were rated on a 7-point Likert scale (\emph{7 = highest plausibility}) and assessed out of VR after each VR exposure. The questions were adapted to address multiple virtual others instead of a singular virtual human.

\subsubsection{Co-Presence (CoP) and Impression of Interaction Possibilities (IIP)}
We assessed co-presence (CoP) and the impression of interaction possibilities (IIP) with the equivalent sub-scales of the questionnaire proposed by \cite{poeschlMeasuringCoPresenceSocial2015}. The seven items were rated on a 7-point Likert scale (\emph{7 = highest CoP or IIP}) and assessed out of VR after each VR exposure.

\subsubsection{Affective Appraisal}
We assessed the virtual other's humanness~(UVI-H) and eeriness~(UVI-E) with the equivalent dimensions of the revised \emph{Uncanny Valley Index}~(UVI)~\citep{hoMeasuringUncannyValley2017}. 
The 14 items were rated using semantic differentials ranging from -3 to 3. The results were mapped to a scale of 1 to 7~(\emph{7~=~highest UVI-H and UVI-E}). The UVI's task introduction was adapted to refer to multiple virtual others instead of a singular virtual character. The items were assessed out of VR after each VR exposure. 

\subsubsection{Spatial Presence (SP)}
We assessed spatial presence~(SP) with the SP sub-dimension of the \emph{Igroup Presence Questionnaire}~(IPQ)~\citep{schubertExperiencePresenceFactor2001}. The five items are rated on a 7-point Likert scale (\emph{6 = highest SP}) and were assessed out of VR after the last VR exposure. We further assessed the \emph{One Item Presence Scale} (OIPS)~\citep{bouchard2004reliability} in VR after each VR exposure. The OIPS comprises one item rated on a 10-point scale (\emph{10 = highest SP}).

\subsubsection{Sense of Embodiment (SoE)}
We assessed the SoE and its dimensions virtual body ownership~(VBO), agency~(AG), and the change~(CH) in the body schema using the \emph{Virtual Embodiment Questionnaire~(VEQ)} \citep{roth2020construction}. Additionally, we measured self-location~(SL) using the items introduced by \cite{fiedler2023selfidentification}.
Each factor is evaluated with four items rated on a 7-point Likert scale~(\emph{7 = highest SoE}) and was assessed out of VR after the last VR exposure. We further assessed one significant item of each dimension in VR after each VR exposure, rated on a 10-point scale (\emph{10~~=~ highest SoE}).

\subsubsection{Self-Identification (SI)}
We assessed self-identification (SI) with eight items concerning self-attribution and self-similarity as introduced by \cite{fiedler2023selfidentification}. The items were rated on a 7-point Likert scale~(\emph{7 = highest SI}), and were assessed out of VR after the last VR exposure. We further assessed one significant item for self-attribution and one for self-similarity in VR after each VR exposure, rated on a 10-point scale~(\emph{10 = highest SI}).

\subsubsection{\added[id=R1]{Control Measures}}
\label{subsubsec:control}
\added[id=R1]{We controlled for interindividual differences between participants by considering their VR experience and perceived signs of simulator sickness. Therefore, we assessed VR experience before the first VR exposure as a categorical variable measuring viewing time in hours, rated in four categories~(\emph{0, 1 to 10, 10 to 50, and over 50 hours}). Furthermore, we assessed simulator sickness before the first and after the last VR exposure using the \emph{Simulator Sickness Questionnaire}~\citep{kennedy1993simulator}. It consists of 32 items capturing symptoms associated with simulator sickness. The total score ranges from 0~to~235.62 (\emph{235.62 = strongest}). We consider the change between the pre and post-VR assessments. An increase in score indicates the occurrence of simulator sickness due to VR usage. }


\subsection{Apparatus}
\label{subsec:apparatus}

\subsubsection{Hard- and Software} 
\label{subsubsec:system}
 We implemented the VR application using Unity in version 2020.3.25f1~\citep{unity2020gameengine}. \deleted[id=R2]{The virtual environment was based on Unity assets, which we adapted to create a rather neutral background, as depicted in \mbox{\autoref{fig:teaser}}}. A Valve Index HMD~\citep{valve2020index} and two Valve Index controllers were integrated using SteamVR~\citep{value2021steamvr} and the corresponding Unity plug-in in version~2.6.1\footnote{\url{https://assetstore.unity.com/packages/tools/integration/steamvr-plugin-32647}}~(see \autoref{fig:methods}, C). The HMD has a display resolution of $1440\times1600$\, pixels per eye, and a total field of view of \SI[parse-numbers=false]{114.1\times109.4}{\degree} \citep{wolf2022observationdistance}. It ran at a refresh rate of \SI{120}{\Hz}. The tracking area ($\SI{3}{\m}\times\SI{4}{\m}$) was set up with four SteamVR Base Stations~2.0. We carefully routed the HMD's cable leading to a high-end workstation~(NVIDIA GeForce RTX 2080 Ti, 32 GB RAM, Intel Core i7-9700K CPU, Windows 10). The participants' fingers were tracked via the controllers' proximity sensors. We did not provide eye and facial expression tracking. For body tracking, we employed the markerless tracking system from Captury \citep{stoll2011motiontracking}~(see \autoref{fig:methods}, D). Movements were tracked at a rate of \SI{100}{\Hz} with eight FLIR Blackfly S BFS-PGE-16S2C RGB cameras connected to a high-end workstation~(NVIDIA GeForce RTX 3080 Ti, 32 GB RAM, AMD Ryzen 9 5900x, Ubuntu 20.04.6 LTS) using two 4-port 1 GBit/s ethernet frame-grabber. The workstation ran Captury Live in version 254b~\citep{captury2021captruylive} streaming the body pose to the VR system integrating Captury's Unity plug-in\footnote{\url{https://captury.com/resources/}}. 
Questionnaires were assessed on a desktop PC using LimeSurvey 4.5~\citep{limeservey2020fourFive}.

\subsubsection{\added[id=R2]{Virtual Environment}}
\begin{figure*}[!t]
  \centering
  \includegraphics[width=\linewidth]{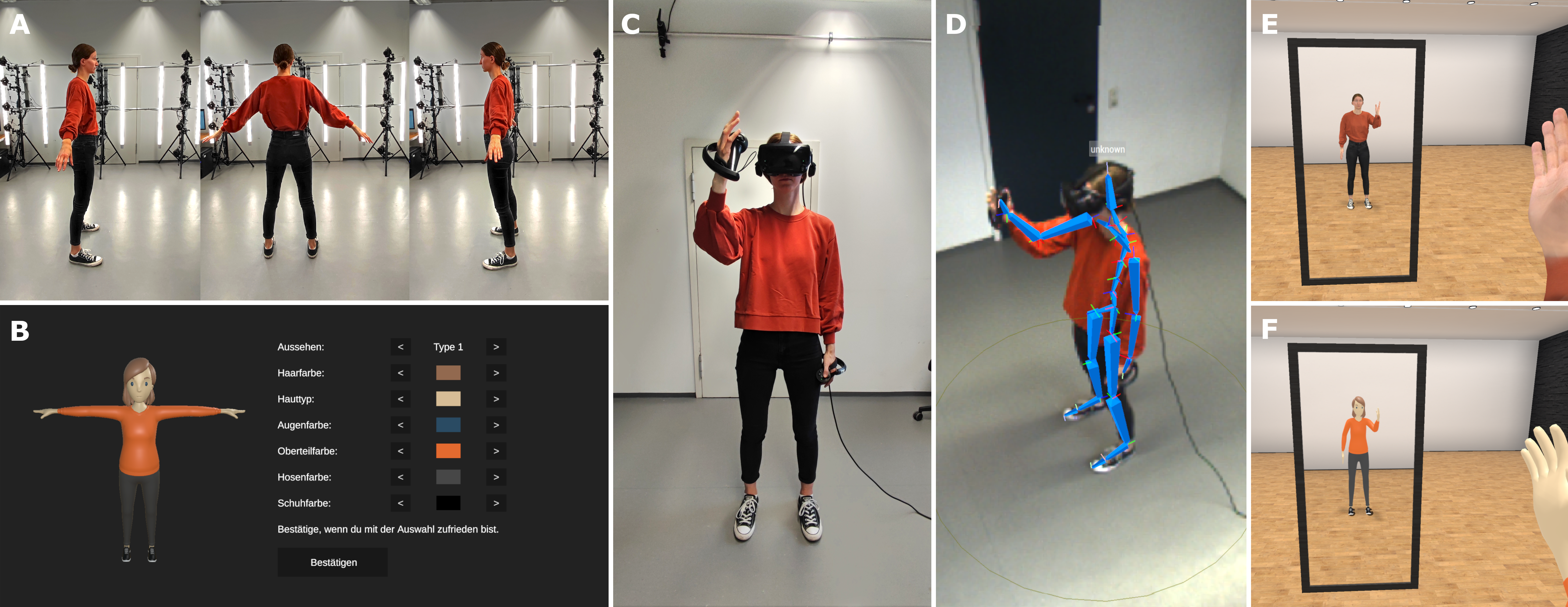}
  \caption{The scan process for personalization of realistic avatars (A), the GUI for customization of stylized avatars (B),  an immersed person~(C), Captury Live's tracking view~(D), and the first- and third-person perspective of an embodied realistic (E) and a stylized (F) avatar in the~mirror. }
  \label{fig:methods}
\end{figure*}

\added[id=R2]{The virtual environment featured a wooden floor enclosed by white walls accented with black elements, creating a rather neutral and minimalist aesthetic (see \autoref{fig:teaser}). Semi-transparent footprints indicated the participant's starting position and a door to the wall behind the participant aided orientation within the space, representing a potential exit. The spacious virtual environment measured \mbox{\SI[parse-numbers=false]{12\times12}{\meter} }, providing ample room for body movement tasks, movement towards the group of virtual others, and participation in the odd-one-out paradigm within a group of seven virtual humans.  A black framed mirror (\mbox{\SI[parse-numbers=false]{1\times2}{\meter} }) was initially hidden and later on appeared \mbox{\SI{2}{\meter}} in front of the participant (see \autoref{fig:methods}, E and F).} \looseness=-1

\subsubsection{Creation of Realistic Virtual Humans}
\label{subsubsec:realistic}
We employed a custom-made photogrammetry rig to scan participants in a laboratory at the University of Würzburg. The rig featured 15 Canon EOS 2000D cameras equipped with high-quality \SI{35}{mm} wide angle lenses\footnote{\url{https://www.canon.de/lenses/ef-35mm-f-2-is-usm-lens/}} arranged in a $5 \times 3$ grid~(see \autoref{fig:methods}, A). The small number of cameras and the streamlined design ensured unhindered accessibility to the scanning process, even within the confines of a small laboratory. Six LED tubes were used to light the subjects uniformly. 
To capture the participants from every angle, they were scanned from the front, back, left, and right. An additional scan captured the empty background, which was used for background subtraction in the photogrammetry step. The images were automatically processed using the commercial photogrammetry software Agisoft Metashape~\citep{metashape}, resulting in four dense point clouds, each defining a partial scan of the subject. 

For generating realistic virtual humans from this input data, we use a template-fitting approach based on the work of \cite{achenbachFastGenerationRealistic2017}. They used a photogrammetry rig to generate a single dense point cloud of the subject, to which an animatable statistical human body model was fitted. Non-rigid ICP~\citep{bouaziz2014registration-tutorial} and fine-scale surface deformation were employed to optimize the statistical template model's alignment, pose, and shape. 

With our approach, participants turned $90^{\circ}$ for each of the scans, and the four point clouds depict the subjects in slightly varying poses. We adopted the avatar generation method of \cite{achenbachFastGenerationRealistic2017} accordingly. On each point cloud, we selected $k$ landmarks, which served as initial correspondences between the template model and the partial scans ($k \in \{23, 9, 6, 6\}$ for the front, back, left, and right scans, respectively). Pose parameters of the template model were then individually optimized for each of the partial scans, while the shape parameters were jointly optimized such that the resulting shape matched all of the point clouds. Similarly, the fine-scale deformation was reformulated to minimize the shape difference between the template model and all of the partial scans simultaneously. Texture information for the resulting geometry was then generated by following the method of \cite{wenningerRealisticVirtualHumans2020}, which generated partial textures from the camera calibration data resulting from the photogrammetry step. Partial textures were then stitched together via a graph-cut based optimization and Poisson Image Editing~\citep{Boykov2001_GraphCut, Perez2003PoissonImageEditing}. \looseness=-1

\subsubsection{Creation of Stylized Virtual Humans}
\label{subsubsec:individualized}
The stylized virtual humans are based on a female\footnote{\url{https://www.cgtrader.com/3d-models/character/woman/cartoongirl009-girl}} and a male\footnote{\url{https://www.cgtrader.com/3d-models/character/man/cartoonman035-man}} character 3D model in a cartoon style, both wearing long-sleeved clothing. The models were rigged using Mixamo~\citep{mixamo2023}. Additional hairstyles were created using Blender in version 2.8.0~\citep{blender2019}. To create individualized stylized self-avatars, participants were provided an array of customization options using a graphical user interface implemented using Unity~\citep{unity2020gameengine} as depicted in \autoref{fig:methods},~B. The options included three short and two long hairstyles, ten skin colors based on the Monk Skin Tone Scale~\citep{monk_2019}, thirteen eye colors with variations of blue, green, brown, grey, and black, as well as thirteen hair colors, ranging from light gray, blond, red, and~brown to black. A selection of 33 colors for clothing and shoes was provided. Participants were instructed to choose a configuration that looked similar to them. \looseness=-1

\subsubsection{Implementation of the Self-Avatar}
Participants could observe their self-avatar's virtual body from the first-person and third-person perspectives in a virtual mirror (see \autoref{fig:methods}, E and F). We implemented embodiment by retargeting Captury's tracked body pose to the avatar in real-time. We addressed inaccuracies in the end-effectors' positions caused by variations in skeletal structure and segment lengths between the target avatar and the source pose following the approach of \cite{wolfExploringPresenceAvatar2022}. Thereby, we aligned the avatar's end-effectors (i.e., head, hands, and feet) with the tracked end-effectors of the user implementing an IK-supported pose optimization step using FinalIK in version 2.1~\citep{rootmotion2019finalik}. The system's motion-to-photon latency for full embodiment averaged \SI{35}{\ms} for hand and \SI{116}{\ms} for other body movements. It was assessed by frame counting~\citep{stauffert2020latencyreview}, tracking the time difference between real-world movements and the rendered corresponding avatar movements. 
Stylized self-avatars were automatically scaled to the participant's eye height, while the personalized scan process implicitly predetermined the height of realistic self-avatars.\looseness=-1

\subsubsection{Implementation of the Group of Virtual Others}
\label{subsec:virtualOthers}
The group of virtual others consisted of six virtual humans with three possible configurations of appearance. The virtual humans were either (1)~all realistic, (2)~all stylized, or (3)~evenly distributed between stylized and realistic. The virtual humans' genders were distributed evenly between males and females. All virtual humans were positioned in a circular pattern with a radius of ~\SI{2.5}{\meter}, facing the group's center. There was one additional empty spot for the participant to join the group (see \autoref{fig:teaser}).

\paragraph{Appearance of Group Members}
We generated six realistic virtual humans by scanning three male and three female volunteers dressed in casual attire in a laboratory at the University of Würzburg. 
The volunteers' ages ranged from 21~to~25~$(M~=~23.33, SD~=~1.86)$,  they gave explicit consent for using their virtual reconstructions in a scientific context, and they were not compensated for participation. Three of the resulting realistic virtual humans are depicted in \autoref{fig:teaser}. The stylized virtual others were created using random configurations of the customization options described in \autoref{subsubsec:individualized}. Therefore, we excluded the base avatar type that participants initially chose for their customized stylized self-avatar to avoid unintentionally creating a digital twin of the participant. Stylized virtual others were scaled to a body height of~\SI{165.8}{\cm} for females and \SI{178.9}{\cm} for males, relating to the body height average of the German population in 2021\footnote{\url{https://www.destatis.de/DE/Themen/Gesellschaft-Umwelt/Gesundheit/Gesundheitszustand-Relevantes-Verhalten/Tabellen/liste-koerpermasse.html}}.

\paragraph{Animation and Behavior of Group Members}

\added[id=R2]{To avoid potential bias caused by different emotional dispositions of the virtual others \mbox{\citep{volonteEffectsInteractingCrowd2020}}, we aimed to keep animation and behavior constant between the groups' configurations. Therefore, the following description of animation and behavior equally applies to both types of virtual humans.}
The virtual human's body animation was drawn from eight distinct idle animations exported from Mixamo~\citep{mixamo2023}. To simulate the input behavior of virtual others, we additionally captured nine pointing animations using an Xsens MVN Link\footnote{\url{https://www.movella.com/products/motion-capture/xsens-mvn-link}} motion capture suit with Xsens MVN Record in version 2022.0.0~\citep{xsens2022}. Two of the pointing gestures were performed with the left hand and seven with the right hand. All animations maintained fixed world positions and rotations, ensuring the virtual humans remained stationary, facing the group's center. No facial expressions were displayed. \looseness=-1

The virtual others' behavior followed a rule-based, event-driven approach. Accordingly, each virtual human's behavior adhered to a currently assigned state, determining its body animation and target of attention, i.e., where they looked at in the virtual environment. All transitions between states aligned with the ongoing instructions provided to the participants, thereby yielding credible behavior that mirrored the study's procedural structure.
In the default state, the virtual humans would either remain inactive or look around, performing idle animations, thereby gazing at the group center or occasionally shifting their target of attention to the eyes of other virtual humans or the participant. In an introductory state, the virtual humans' target of attention switched to the participant's eyes once they came closer than 3.5 meters to the group center, which was forced when the participant was told to join the group. All virtual others then briefly looked at the participants to simulate that they recognized them. 
Once the logic tasks began, the virtual humans entered an operational state, directing their attention to the task objects or the input tablet respectively. Upon the participant being prompted to select an object, the virtual humans shifted state, eliciting a pointing animation towards a random solution input on the tablet.

\paragraph{Framing the Group of Virtual Others}
\label{par:framing}
There was no explicit framing regarding whether the virtual others were human or computer-controlled. In all questionnaires, they were referred to as \textquote{virtual characters}, deliberately avoiding implications of humanness, which the term \textquote{virtual humans} might have. In VR, virtual others were simply referred to as \textquote{others}, and participants were instructed not to communicate with them, as each individual was required to solve the logic tasks independently. This also suggested participants not expect virtual others to initiate conversation. It was implied that the others were engaged in solving the logic tasks as well, as the audio instructions, presented while the others were also displayed, addressed everyone collectively, for example, starting with \textquote{Nice to have you all here. Today, you will each solve brainteasers in multiple rounds of the game}.\looseness=-1

\subsection{Experimental Tasks}
\label{subsec:procedure}

\subsubsection{Body Movement Tasks}
\label{subsubsec:embodimenttasks}
Participants engaged in three body movement tasks (i.e., moving the fingers, waving the hand, and walking in place) in front of a virtual mirror for~\SI{20}{\second}~each. The tasks were adapted from prior work~\citep{wolfBodyWeightPerception2020,waltemateImpactAvatarPersonalization2018} aiming to familiarize participants with the virtual body and to induce an SoE by establishing visuomotor coherence~\citep{slaterInducingIllusoryOwnership2009} from both the first and third-person perspectives. \looseness=-1

\subsubsection{Logic Tasks}
\label{subssec:logictask}
We implemented a VR Odd-One-Out paradigm, wherein participants for each task were presented with a set of five primitive 3D objects and were required to identify the unique \enquote{odd} item within the set. We chose this type of logic task as it is a singular, non-cooperative, and non-verbal task that can be replicated in a repeated measures design by slightly changing single primitive characteristics, still requesting comparable cognitive demand for each repeated measure. 
Each VR exposure comprised five Odd-One-Out tasks designed with increasing difficulty \added[id=R1]{to keep participants' concentration and attention level high while avoiding boredom}. Initially, participants solved three easy tasks wherein the odd object differed in a single property, such as color, size, or geometry. Subsequently, one intermediate task challenged participants to identify an object differing in two of the named properties, and finally, one advanced task demanded participants to discern the odd object solely based on its shape, considering factors like edges or elongation (see \autoref{fig:teaser}). For easy tasks, objects were shown for 15 seconds, and for intermediate and advanced tasks, objects were presented for 25 seconds.
The tasks were created based on six different geometries (cube, cuboid, sphere, cylinder, capsule, pyramid), six distinct colors 
, and three sizes. 
Upon completion of the object presentation phase, the objects disappeared, and a tablet for solution input was displayed for a total of 20 seconds. Participants could click one out of five buttons, each representing one of the objects shown. We, therefore, implemented small box colliders on the self-avatars' index distal phalanges, allowing the participants to interact with the tablet's buttons through their virtual bodies. \looseness=-1

\begin{figure}[!th]
 \centering
 \includegraphics[width=0.5\columnwidth,trim={0mm 0mm 0mm 0mm},clip]{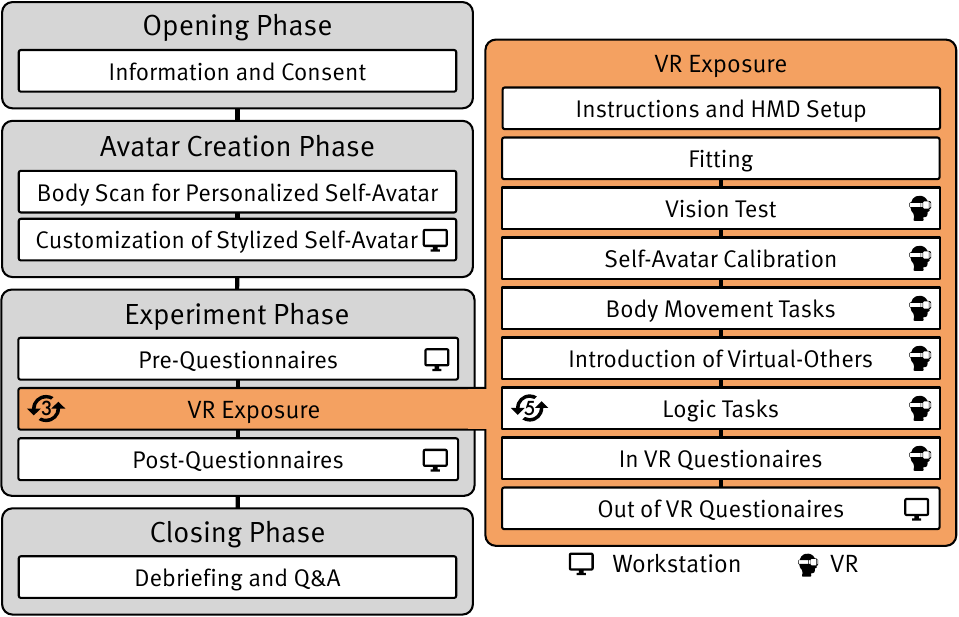}
 \caption{
 Overview of the experimental procedure (left) and the repeated VR exposures (right). All participants underwent the same procedure regardless of their randomly assigned test condition. The icons on the right side of the steps indicate whether the participants used a workstation or were immersed in VR. The icons on the left side of the steps indicate the amount of repetition of the respective step. The VR exposure was repeated for each group configuration, i.e., the within-subject factor.  All logic tasks are defined in \autoref{subssec:logictask}. The overall procedure is described in detail in \autoref{subsec:procedure}.}
 \label{fig:procedure}
\end{figure}

\subsection{Procedure}
The study took place in a laboratory at the University of Würzburg and followed a controlled experimental procedure that took around \SI{95}{\minute}~($M=94.83$, $SD=11.79$). All participants underwent the exact same procedure regardless of their randomly assigned test condition. This includes the creation of both a personalized and a customized self-avatar. The procedure was divided into four phases and is visualized in \autoref{fig:procedure}. \looseness=-1

\subsubsection{The Opening Phase and the Self-Avatar Creation Phase}
In the opening phase, participants first read the study information and gave explicit written consent for participation. In the following avatar creation phase, participants were guided to an adjacent room,  where the body scan was performed as described in~\autoref{subsubsec:realistic}. Study data and body scan data were pseudonymized separately to avoid de-anonymization. Back in the main laboratory, the participants customized their stylized self-avatar as described in~\autoref{subsubsec:individualized}, while the experimenter finalized the personalized self-avatar. 

\subsubsection{The Experiment Phase}
In the experiment phase, participants answered pre-questionnaires on demographic data before three consecutive VR exposures took place. For each of the three VR exposures, the experimenter demonstrated how to wear the HMD correctly and instructed participants on how to adjust the interpupillary distance to their eyes and the integrated headphones to their ears. Once immersed in the virtual environment, participants were guided via prerecorded audio instructions, following a linear procedure. Participants underwent a short vision test, and the self-avatar calibration process was conducted. Subsequently, a virtual mirror was introduced and shown in front of the participants, who then engaged in three brief body movement tasks~(see~\autoref{subsubsec:embodimenttasks}). After the tasks, the mirror disappeared, and the participants were introduced to the virtual others. The virtual others~(see~\autoref{subsec:virtualOthers}) then appeared two meters in front of the participants, who were then told to walk towards the group, becoming a part of it. After a detailed introduction, participants familiarized themselves with the tablet's functionality, and a set of five consecutive logic tasks were applied one by one~(see~\autoref{subssec:logictask}). After the last logic task, the virtual others disappeared, and the participants answered seven short questions verbally in VR on spatial presence, the SoE, and self-identification. Participants then took off the HMD and answered questionnaires on the perception of the virtual others' plausibility, co-presence, and affective appraisal in the anteroom. After all VR exposures, participants answered post-questionnaires out of VR on spatial presence, the SoE, and self-identification. See~\autoref{subsec:measures} for an overview of all measures. 

\subsubsection{The Closing Phase}
Lastly, the experimenter debriefed participants, named manipulated variables, and answered all questions.

\subsection{Statistical Analysis}
\label{subsec:statistical-analysis}
Statistical analyses were performed with R in version 4.3.0~\citep{RCore} and conducted at a significance level of $\alpha=.05$. \added[id=R1]{For all models, we tested whether our control measures (\autoref{subsubsec:control} significantly impacted any dependent variables and ensured they were independent of the predictor variables. If applicable, we incorporated the respective control measures into the models described below.}

We calculated mixed ANOVAs with self-avatar (between-subject) and virtual others~(within-subject) as predictors and one of the dependent measures assessed after each VR exposure as the outcome variable using the \emph{rstatix} package \citep{rstatix_R}. Effect sizes were determined using generalized eta-squared, $\eta^2_G$~\citep{bakemanRecommendedEffectSize2005}.
If the assumption of sphericity was violated, we reported Greenhouse-Geisser correction values. In cases of significant within-subject effects, we calculated pairwise post-hoc comparisons with Bonferroni-adjusted p-values. For significant interaction effects, we performed pairwise post hoc comparisons for each between-subject factor, also using Bonferroni-adjusted p-values.
Furthermore, we calculated Welch's t-tests for independent samples and Cohen's $d$ effect sizes to examine the effect of self-avatar on the measures assessed once after the last VR exposure.

\section{Results}
\label{sec:results}

\subsection{Perception of Virtual Others}

\subsubsection{Appearance and Behavior Plausibility (ABP)}
Self-avatar, $F(1,46)=.198,\, p=.658,\ \eta^2_G=.002$, virtual others, $F(2,92)=2.60,\, p=.08,\ \eta^2_G=.024$, and the interaction between self-avatar and virtual others, $F(2,92)=.050,\, p=.951,\ \eta^2_G=.000$, had no significant effect on ABP.

\begin{table}[!t]
   \tiny
  \caption{Descriptive statistics of our measures assessed after each VR exposure compared for each self-avatar and the three configurations of virtual others.}  
  \label{tab:descriptiveStatistics}
  \centering
\begin{tabularx}{\textwidth}{X@{\hspace{0pt}}c@{\hspace{4pt}}c@{\hspace{5pt}}c@{\hspace{7pt}}c@{\hspace{5pt}}c@{\hspace{7pt}}c@{\hspace{5pt}}c@{\hspace{5pt}}}
    \toprule
              & &  \multicolumn{2}{c}{Stylized Others}   & \multicolumn{2}{c}{Mixed Others}  & \multicolumn{2}{c}{Realistic Others}   \\
    
     \cmidrule(l{1pt}r{7pt}){3-4} \cmidrule(l{1pt}r{7pt}){5-6} \cmidrule(l{1pt}r{7pt}){7-8}
      \multicolumn{2}{c}{\textbf{}}   &   Stylized Self & Realistic Self  & Stylized Self & Realistic Self    & Stylized Self & Realistic Self\\ 
      \textbf{Measures} &   Range & $M$ ($SD$) & $M$ ($SD$) & $M$ ($SD$) &  $M$ ($SD$) &  $M$ ($SD$) &  $M$ ($SD$) \\
    \midrule
        
    \textbf{Virtual Human Plausibility}   \\
        \hspace{1mm} Appearance \& Behavior   & [1 -- 7]          & $4.73$ ($1.29$)    &  $4.56$ ($1.30$)   & $4.55$ ($0.95$)  & $ 4.49$ ($1.17$) &  $4.99$ ($0.99$)    & $4.89$ ($1.15$)   \\
        \hspace{1mm} Match with VE  & [1 -- 7]     & $4.79$ ($1.43$)    &  $4.25$ ($1.60$)   & $4.24$ ($1.30$)  & $4.46$ ($1.52$) &  $5.09$ ($1.37$)    & $5.27$ ($1.15$) \vspace{0.5em}\\
    \textbf{Co-Presence}                   \\
        \hspace{1mm} Co-Presence    & [1 -- 7]          & $4.29$ ($1.49$)   &  $3.58$ ($1.44$)   & $4.97$ ($  1.29$)  & $ 4.50$ ($  1.21$) &  $5.38 $ ($ 1.15$)    & $5.03 $ ($ 1.24$)   \\
        \hspace{1mm} Interaction Possibilities & [1 -- 7]             & $2.28$ ($1.13$)    &  $1.97$ ($0.96$)   & $2.73$ ($1.17$)  & $2.70$ ($1.12$) &  $3.25$ ($1.51$)    & $ 3.06$ ($1.68$) \vspace{0.5em}\\
    \textbf{Affective Appraisal }      \\
        \hspace{1mm} Humanness      & [1 -- 7]  & $2.49$ ($0.96$)    &  $2.23$ ($0.95$)   & $3.37$ ($0.95$)  & $3.57$ ($1.26$) &  $4.28$ ($1.37$)    & $4.23$ ($1.19$)   \\
        \hspace{1mm} Eeriness       & [1 -- 7]          & $2.99$ ($1.05$)    &  $3.34$ ($1.37$)   & $3.30$ ($1.01$)  & $3.67$ ($1.05$) &  $3.48$ ($1.00$)    & $3.33$ ($0.81$) \vspace{0.5em}\\

\textbf{Sense of Embodiment}    \\
    \hspace{1mm} Virtual Body Ownership & [1 -- 10] & $5.08$ ($2.12$)    &  $6.21$ ($1.74$)   & $4.75$ ($2.29$)  & $6.38$ ($1.84$) &  $4.50$ ($2.30$)    & $6.54$ ($1.79$)   \\
    \hspace{1mm} Agency          & [1 -- 10] & $6.29$ ($1.88$)    &  $6.79$ ($1.82$)   & $6.08$ ($2.08$)  & $6.79$ ($1.47$) &  $5.83$ ($1.79$)    & $7.00$ ($1.56$) \\ 
    \hspace{1mm} Change          & [1 -- 10] & $4.04$ ($3.38$)    &  $4.50$ ($2.72$)   & $3.92$ ($3.31$)  & $4.67$ ($2.73$) &  $3.71$ ($3.04$)    & $4.71$ ($2.69$) \\ 
    \hspace{1mm} Self-Location   & [1 -- 10] & $5.83$ ($2.08$)    &  $5.42$ ($2.26$)   & $5.21$ ($2.36$)  & $5.62$ ($2.14$) &  $4.83$ ($2.60$)     & $5.67$ ($2.12$) \vspace{0.5em}\\                                        
    \textbf{Self-Identification}  \\
    \hspace{1mm} Self-Identification  &  [1 -- 10]         & $5.02$ ($2.03$)    &  $6.75$ ($1.68$)   & $4.48$ ($1.90$)  & $7.08$ ($1.04$) &  $3.96$ ($1.77$)     & $6.96$ ($1.46$)\vspace{0.5em}\\ 

     \textbf{Spatial Presence}       \\    
    \hspace{1mm} Spatial Presence    &   [1 -- 10]         & $6.46$ ($2.00$)    &  $6.33$ ($1.27$)   & $6.42$ ($1.44$)  & $6.88$ ($1.48$) &  $6.33$ ($2.12$)     & $7.00$ ($1.47$)\\  
       
   \bottomrule
\end{tabularx}
\end{table}

\subsubsection{Match with Virtual Environment (MVE) }
Virtual others had a significant main effect on MVE, $F(1.71,78.88)=5.198,\, p=.011,\ \eta^2_G=.064$.  On average, realistic virtual others were rated significantly higher on MVE compared to mixed virtual others, $F(47)=3.65,\, p_{adj}=.002$, but not compared to stylized virtual others,  $F(47)=2.05,\, p_{adj}=.139$. Furthermore, self-avatar, $F(1,46)=.036,\, p=.85,\ \eta^2_G=.000$, and the interaction between self-avatar and virtual others, $F(1.71,78.88)=1.227,\, p=.294,\ \eta^2_G=.016$, had no significant effect on MVE.

\subsubsection{Co-Presence}
Virtual others had a significant main effect on co-presence, $F(1.72,79.05)=29.544,\, p<.001,\ \eta^2_G=.142$. On average, realistic virtual others were rated significantly higher on co-presence than mixed virtual others, $F(47)=3.06,\, p_{adj}=.011$, and mixed virtual others were rated significantly higher on co-presence compared to stylized virtual others, $F(47)=5.57,\, p_{adj}<.001$. Furthermore, self-avatar, $F(1,46)=2.45,\, p=.124,\ \eta^2_G=.038$, and the interaction between self-avatar and virtual others had no significant effect on co-presence, $F(1.72,79.05)=0.608,\, p=.523,\ \eta^2_G=.003$. 

\subsubsection{Impression of Interaction Possibilities (IIP)}
Virtual others had a significant main effect on IIP, $F(1.66,76.47)=21.973,\, p<.001,\ \eta^2_G=.101$. On average, 
realistic virtual others were rated significantly higher on IIP than mixed virtual others, $F(47)=3.37,\, p_{adj}=.005$, and mixed virtual others were rated significantly higher on IIP compared to stylized virtual others, $F(47)=4.13,\, p_{adj}<.001$. Furthermore, self-avatar, $F(1,46)=0.298,\, p=.588,\ \eta^2_G=.005$, and the interaction between self-avatar and virtual others, $F(1.66,76.47)=0.408,\, p=.629,\ \eta^2_G=.002$, had no significant effect on IIP.

\subsubsection{Humanness}
Virtual others had a significant main effect on humanness, $F(1.66,76.14)=67.911,\, p<.001,\ \eta^2_G=.331$.  On average, realistic virtual others were rated significantly higher on humanness compared to mixed virtual others, $F(47)=4.76,\, p_{adj}<.001$,  and mixed virtual others were rated significantly higher compared to stylized virtual others,  $F(47)=8.62,\, p_{adj}<.001$. Furthermore, self-avatar, $F(1,46)=.022,\, p=.884,\ \eta^2_G<.001$, and the interaction between self-avatar and virtual others had no significant effect on humanness, $F(1.66,76.14)=.993,\, p=.362,\ \eta^2_G=.007$. 

\begin{figure}[!t]
  \centering
  \includegraphics[width=\linewidth]{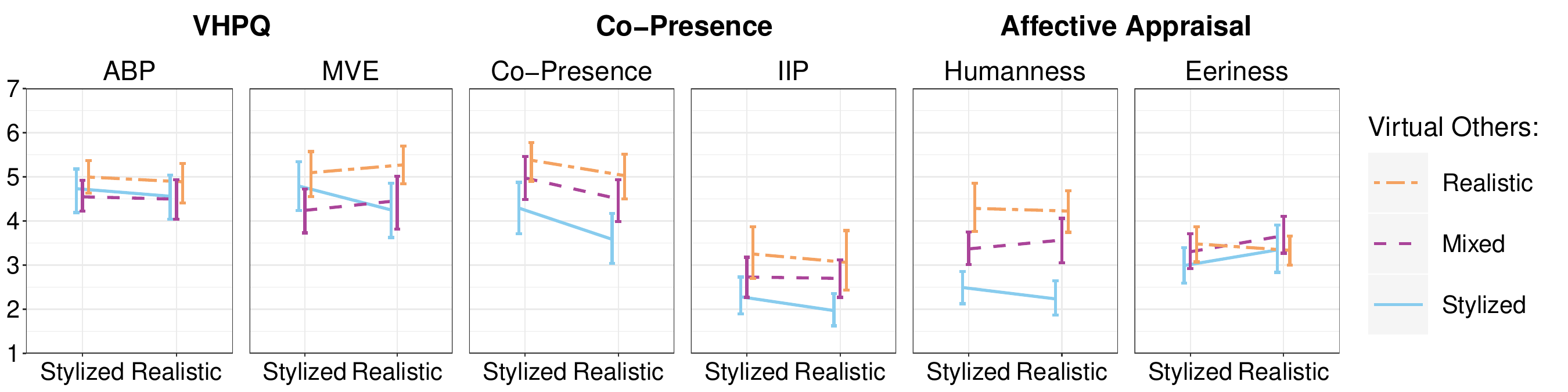}
  \caption{The interaction plots show the type of self-avatar (x-axis) and the contrast between realistic, mixed, and stylized virtual others for the evaluation of the perception of virtual others. Error bars represent 95 \% confidence intervals estimated using bootstrapped standard deviations.}
  \label{fig:interaction_others}
\end{figure}

\subsubsection{Eeriness}
Self-avatar, $F(1,46)=.661,\, p=.420,\ \eta^2_G=.008$, virtual others, $F(1.74,79.88)=1.878,\, p=.165,\ \eta^2_G=.017$, and the interaction between self-avatar and virtual others, $F(1.74,79.88)=1.458,\, p=.239,\ \eta^2_G=.013$, had no significant effect on eeriness.

\subsection{Perception of Self-Representation}

\subsubsection{Virtual Body Ownership (VBO)}
Self-avatar had a significant main effect on VBO assessed in VR after each VR exposure, $F(1,46)= 8.802,\, p=.005,\ \eta^2_G=.139$. On average, realistic self-avatars~($M = 6.38,\ SD = 1.79$) were rated higher on VBO than stylized self avatars~($M = 4.78,\ SD = 2.24$). Virtual others, $F(1.76,81.01)=0.204,\, p=.788,\ \eta^2_G=.001$, and the interaction between self-avatar and virtual others had no significant effect on VBO in VR,  $F(1.76,81.01)=2.646,\, p=.084,\ \eta^2_G=.009$.

Furthermore, self-avatar had a significant effect on VBO assessed out of VR after the last VR exposure, $t(45.00) = 2.97,\ p = .005,\ d = .86$. On average, realistic self-avatars~($M = 4.68,\ SD = 1.05$) were rated higher on VBO than stylized self-avatars ~($M = 3.70,\ SD = 1.23$). 

\subsubsection{Agency}
Self-avatar, $F(1,46) = 2.786,\, p=.102,\ \eta^2_G=.049$, virtual others, $F(2,92)=0.312,\, p=.733,\ \eta^2_G=.001$, and the interaction between self-avatar and virtual others had no significant effect on agency assessed in VR after each VR exposure, $F(2,92)=2.024,\, p=.138,\ \eta^2_G=.006$.  

Furthermore, self-avatar had no significant effect on agency assessed out of VR after the last VR exposure, $t(45.56) = 0.249,\ p = .804,\ d = .07$.

\subsubsection{Change}
Self-avatar, $F(1,46) = 0.754,\, p=.390,\ \eta^2_G=.016$, virtual others, $F(2,92)=.178,\, p=.837,\ \eta^2_G<.001$, and the interaction between self-avatar and virtual others had no significant effect on change assessed in VR after each VR exposure, $F(2,92)=1.737,\, p=.182,\ \eta^2_G=.001$. 

Furthermore, \replaced[id=R1]{the covariate VR experience was significantly related to change assessed out of VR after the last VR exposure, $F(1,45) = 8.37,\, p=.006,\ r=.40$. S}{s}elf-avatar had no significant effect on change assessed out of VR after the last VR exposure,  \replaced[id=R1]{$F(1,45) = .52,\, p=.473,\ r=.11$}{$t(38.40) = 0.671, p = .506, d = 0.19$.}

\subsubsection{Self-Location}
Self-avatar and virtual others had a significant interaction effect on self-location assessed in VR after each VR exposure, $F(2,92)=3.145,\, p=.048,\ \eta^2_G=.014$. For stylized self-avatars, realistic virtual others led to significantly lower ratings on self-location than stylized virtual others, $F(23)=-2.85,\, p_{adj}=.027$, and for realistic self-avatars, virtual others had no significant effect on self-location (all $p_{adj} > 0.99$).
Moreover, self-avatar, $F(1,46) = 0.225,\, p=.637,\ \eta^2_G=.004$, and virtual others, $F(2,92)=1.096,\, p=.338,\ \eta^2_G=.005$, had no significant main effect on self-location in VR. 

Furthermore, self-avatar had a significant effect on self-location assessed out of VR after the last VR exposure, $t(45.37) = 3.107,\ p = .003,\ d = 0.90$. On average, realistic self-avatars~($M= 4.32,\ SD=0.96$)  were rated higher on self-location than stylized self-avatars~($M = 3.41,\ SD = 1.08$). See \autoref{tab:descriptiveStatistics_after} for descriptive statistics.

\begin{figure}[!t]
  \centering
  \includegraphics[width=\linewidth]{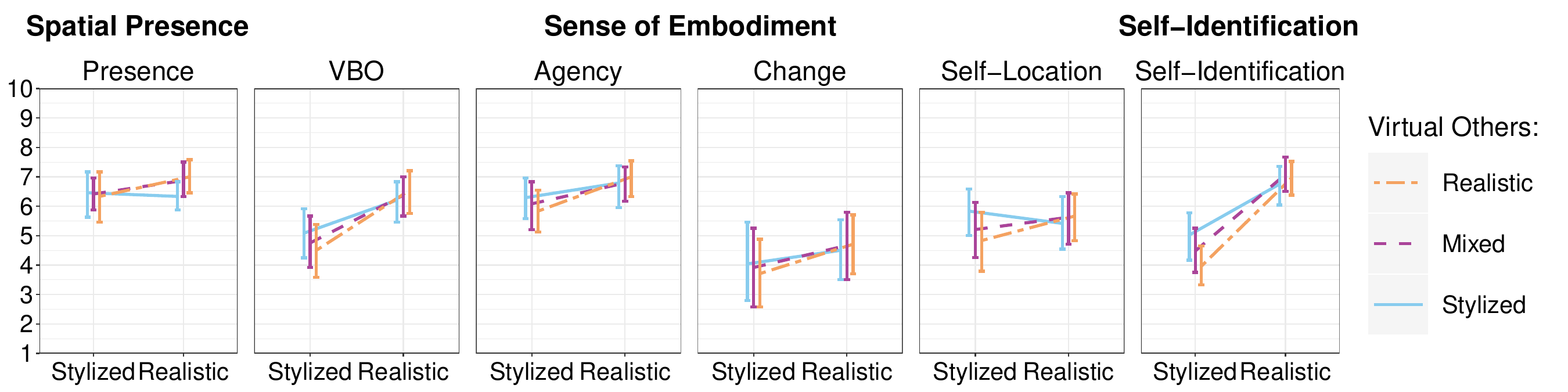}
  \caption{Interaction plots show the type of self-avatar (x-axis) and the contrast between realistic, mixed, and stylized virtual others for the evaluation of spatial presence and the perception of the self-representation. Error bars represent 95 \% confidence intervals estimated using bootstrapped standard deviations.}
  \label{fig:Results}
\end{figure}

\begin{table}[!b]
  \small
  \caption{Descriptive statistics for measures assessed out of VR after the last VR exposure for stylized and realistic self-avatars. All measures' ranges are [1 -- 7].}
  \label{tab:descriptiveStatistics_after}
  \centering
  \begin{tabu} to \columnwidth {X[1.17,l,m]@{\hspace{8pt}}X[0.89,l,m]@{\hspace{8pt}}c@{\hspace{8pt}}c@{\hspace{2pt}}}
    \toprule
     &               & Stylized Self & Realistic Self \\  
   \textbf{Measures}  &               & $M$ ($SD$) & $M$ ($SD$) \\  
    \midrule
    \textbf{Sens of Embodiment} & VBO               &  3.70 (1.23)  & 4.68 (1.05) \\
                                & Agency            &  5.75 (0.76)  & 5.80 (0.69) \\
                                & Change            &  2.97 (1.81)  & 3.26 (1.12) \\
                                & Self-Location     &  3.41 (1.08)  & 4.32 (0.96)  \vspace{0.5em}\\
    \textbf{Self-Identification}
                                & Self-Identification   &  3.41 (1.02)  & 5.07 (0.93)  \vspace{0.5em}\\
    \textbf{Spatial Presence}   & Spatial Presence  &  3.51 (0.89)  & 3.16 (0.74)\\  
   
  \bottomrule
\end{tabu}
\end{table}

\subsubsection{Self-Identification (SI)}
Self-avatar had a significant main effect on SI assessed in VR after each VR exposure, $F(1,46)= 29.149,\, p<.001,\ \eta^2_G=.344$. On average realistic self-avatars~($M = 6.93,\ SD = 1.51$) were rated higher on SI than stylized self avatars~($M = 4.49,\ SD = 1.90$). Also, the interaction between self-avatar and virtual others had a significant effect on SI in VR, $F(1.54,70.66)=6.645,\, p<.005,\ \eta^2_G=.024$. For stylized self-avatars, stylized virtual others led to significantly higher ratings in SI compared to mixed virtual others, $F(23)=2.99,\, p_{adj}=.02$, and realistic virtual others, $F(23)=3.65,\, p_{adj}=.004$. For realistic self-avatars, virtual others had no significant effect on SI (all $p_{adj} > .777$). 

Furthermore, self-avatar had a significant effect on SI assessed out of VR after the last VR exposure, $t(45.54) = -5.90,\ p < .001,\ d = -1.70$. On average, realistic self-avatars~($M = 5.07,\ SD = 0.93$) were rated higher on SI than stylized self-avatars~($M = 3.41,\ SD = 1.02$).  See \autoref{tab:descriptiveStatistics_after} for descriptive statistics.

\subsection{Spatial Presence}
There were no significant effects of self-avatar, $F(1,46) = 0.627,\, p=.432,\ \eta^2_G=.010$, virtual others, $F(2,92)=1.143,\, p=.323,\ \eta^2_G=.006$, and the interaction between self-avatar and virtual others, $F(2,92)=2.118,\, p=.126,\ \eta^2_G=.010$ on spatial presence assessed in VR after each VR exposure.  

Furthermore, self-avatar had no significant effect on spatial presence assessed out of VR after the last VR exposure, $t(44.453) = -1.479,\ p = .146,\ d = -.472$.

\begin{table}[!htb]
\small
\centering
\caption{ \replaced[id=R2]{All hypotheses and explorative evaluations as stated in  \mbox{\autoref{sec:related-work}}}{Hypotheses and results. Original hypotheses from \mbox{\autoref{sec:related-work}} were abbreviated for conciseness}.\looseness=-1}
\label{table:hypos}
\begin{tabularx}{\textwidth}{l@{\hspace{6pt}} X c}
\toprule
      
  \multicolumn{2}{l}{Hypotheses} &  Result \\
\midrule
\multicolumn{2}{l}{\textbf{Virtual Others}}   \\ 
\hspace{1pt} H1:& \replaced[id=R2]{ The manipulation of virtual others' realism and the self-avatar's realism will lead to significantly higher scores in VHP for configurations of higher congruence. \vspace{0.5em}}{Higher scores in VHP for virtual others of high realism.}   &  Rejected\\

\multicolumn{2}{l}{\textbf{Self-Representation}} \\
\hspace{1pt} H2.1:& \replaced[id=R2]{The manipulation of the self-avatar's congruence with the participant will lead to significantly higher scores in VBO and self-identification for realistic self-avatars.}{Higher scores in VBO for realistic self-avatars.}                                 &  Accepted \\
\hspace{1pt} H2.2:& \replaced[id=R2]{The manipulation of the self-avatar's congruence with the participant will not significantly affect agency, change, and self-location. \vspace{0.5em}}{No effect of the self-avatar on Agency, Change, and Self-location.     }          & Partially Accepted \\

\multicolumn{2}{l}{\textbf{Spatial Presence} }\\
\hspace{1pt} H3:& \replaced[id=R2]{The manipulation of the self-avatar's realism, the virtual others' realism, and their congruence will not significantly affect spatial presence.}{No effect of virtual others, self-avatar, and their (in)congruence on Spatial Presence}     & Accepted \vspace{0.5em}\\

\end{tabularx}

\begin{tabularx}{\textwidth}{l@{\hspace{6pt}} X}
  \toprule
  \multicolumn{2}{l}{\added[id=R2]{Explorative Evaluations}}   \\
 \midrule
  \multicolumn{2}{l}{\added[id=R2]{\textbf{Virtual Others}}} \\
 \hspace{1pt} E1.1:  & \hspace{1pt}\added[id=R2]{Evaluation of the impact of the virtual others' realism, and their congruence  with the self-avatar's realism, on co-presence and the impression of interaction possibilities} \\
 \hspace{1pt} E1.2: & \hspace{1pt}\added[id=R2]{Evaluation of the impact of the virtual others' realism, and their congruence  with the self-avatar's realism, on affective appraisal and the Uncanny Valley effect.} \vspace{0.5em} \\

\multicolumn{2}{l}{ \added[id=R2]{\textbf{Self-Presentation}}} \\
 \hspace{1pt} E2.1: & \hspace{1pt}\added[id=R2]{Evaluation of the impact of the congruence between the self-avatar's realism and virtual other's realism on the perception of self-representation. } \vspace{0.5em}\\



\bottomrule
\end{tabularx}

\end{table}
\section{Discussion}
\label{sec:discussion}

Our study investigated how the realism of self-avatars, a group of co-located virtual humans, and related (in)congruencies affected the users' VR experience regarding the perception of virtual others, self-representation, and spatial presence.  \added[id=R2]{In order to aid comprehension, \autoref{table:hypos} restates all hypotheses and explorative evaluations.} \replaced[id=R2]{T}{In t}he upcoming sections, \deleted[id=R2]{we} will delve into our study's findings, address its limitations, and explore the implications of our research.


\subsection{(In)Congruencies and the Perception of Virtual Others}

\subsubsection{Virtual Human Plausibility}

\cite{mal2022virtual} framed the subjective feeling of how reasonable and believable a virtual human appears to a user as virtual human plausibility\added[id=R2]{~(VHP)} arising from the congruence of its cues, eventually influencing a virtual human's appearance and behavior plausibility~(ABP) and perceived match with the~VE~(MVE). 
\added[id=R2]{We hypothesized the manipulation of virtual others’ realism and the self-avatar’s realism to lead to significantly higher scores in VHP for configurations of higher congruence (H1).}
Contrary to our hypothesis, we found no differences in the perceived plausibility of appearance and behavior among virtual others between realistic, mixed, and stylized configurations\deleted[id=R2]{(H1.1)}, homogeneous and mixed groups\deleted[id=R2]{~(H1.2)}, or conditions with (in)congruent realism between self-avatars and virtual others\deleted[id=R2]{(H1.3)}. Despite our deliberate manipulation of realism introducing \replaced[id=R2]{higher order cognitive}{visual} incongruencies \added[id=R2]{ \mbox{\citep{latoschik2022plausibility}}, as described in \mbox{\autoref{subsubsec:condition}}}, these did not seem to affect participants' subjective perceptions of how reasonable and believable the appearance and behavior of the group of others appeared to them.  Participants demonstrated notable flexibility in accepting varying styles as plausible for a group of virtual humans. \replaced[id=R2]{Participants}{They} may have accommodated incongruencies on a cognitive level based on their habituation to virtual environments simulating alternative realities. 
\added[id=R2]{Since we did not specify whether the virtual others were real humans or agents (see \autoref{par:framing}), the rule-based behavior might have suggested that the virtual others were computer-controlled. Consequently, perceiving the virtual others as agents may have reduced the need for them to have a realistic, human-like appearance to be considered believable in an alternative, virtual reality. However, this is an interpretative approach that suggests the need for further research.}
We also consider that the preliminary VHP questionnaire may not have been sensitive to our manipulation, as subsequent findings clearly show a significant manipulation of \added[id=R2]{how} virtual others  \added[id=R2]{were perceived} based on other qualia. 
\looseness=-1

In terms of the virtual others' match with the virtual environment (MVE), a homogeneous group of realistic virtual others received the highest ratings for MVE and was rated significantly higher than the mixed configuration of virtual others, but not the homogeneous stylized configuration. This overall suggests that the virtual environment better aligned with the appearance of realistic virtual humans. Further analyzing the interaction plot presented in \autoref{fig:interaction_others} for MVE indicates that for realistic self-avatars, the (in)congruent stylized configuration of virtual others received the lowest ratings. Conversely, with stylized self-avatars, the congruent stylized configuration of virtual others was more favorably accepted and rated as fitting better within the virtual environment. 
However, we approach this interpretation cautiously as the interaction between self-avatar and virtual others did not yield significant differences. We suggest future research to delve deeper into the descriptively indicated impact of self-representation on the evaluation of virtual others' match with a virtual environment. \looseness=-1

%
%

\subsubsection{Co-Presence and Impression of Interaction Possibilities (IIP)}
\added[id=R2]{In an exploratory evaluation, we investigated the impact of the virtual others' realism, and their congruence  with the self-avatar's realism, on co-presence and the impression of interaction possibilities (E1.1).} The realism of virtual others had a notable effect on co-presence. Our results indicate that configurations featuring realistic virtual others greatly enhanced the perceived co-presence with the group. Specifically, group configurations with exclusively realistic others received higher ratings than the mixed others, and mixed others were rated higher than the all-stylized configuration. \replaced[id=R1]{T}{In line with H1.6, t}his effect was consistent for both self-avatars, suggesting that the (in)congruence between the realism of the self-avatar and the group of others did not have an impact on the sense of \enquote{being there together} \citep{Schroeder2002}. 
While previous research on the impact of the appearance of virtual humans on co-presence has yielded heterogeneous results in general, our findings align closely with the work of \cite{zibrekDonStandClose2017}. The authors discovered that participants preferred realistic virtual humans over stylized representations. While the named work utilized a single virtual human as an agent, we can confirm these findings in a group setting. Further aligning with \cite{zibrekDonStandClose2017}, our exploratory evaluation of eeriness did not suggest increased feelings of unappealing for any of the styles,  which might be the base for realism in virtual humans to benefit co-presence. \added[id=R1]{Further, the systematic review of \cite{ohSystematicReviewSocial2018} indicated the congruence between realism in appearance and behavior to positively predict co-presence. 
In our work, the virtual others displayed rather realistic body movements either exported from Mixamo~\citep{mixamo2023} or captured with a state-of-the-art motion capture system~\citep{xsens2022}. These movements might have been more congruent with the realistic others than the abstract ones, eventually leading to increased co-presence.}\looseness=-1

Interestingly, participants also reported that their impression of interaction possibilities with virtual others was higher for configurations featuring realistic virtual others, an effect aligning closely with the co-presence trend. Notably, the overall ratings for IIP were rather low, possibly due to the rule-based behavior of the virtual others and the explicit framing that verbal communication with the others was not allowed (see \autoref{subsec:virtualOthers}), both limiting the interaction possibilities. Nonetheless, participants still perceived that realistic virtual humans offered more contingencies for interaction. We assume the impression of co-presence and interaction possibility with virtual others to be affected by their congruence with real-world knowledge (i.e., being realistic), aligning with the impression that (real) humans are capable of engaging in interactions (interaction possibilities) while being in a place together (co-presence). \looseness=-1 

However, our findings contrast \deleted[id=R1]{H1.4 and} the work of \cite{latoschikNotAloneHere2019}, who suggested potential incongruencies with participants' expectations caused them to focus more intensely on the surrounding agents when they had mixed appearances. The key distinction between the cited study and ours is that \cite{latoschikNotAloneHere2019} utilized a passive ambient crowd, representing an environmental surrounding rather than virtual entities with which one might want to interact directly. From these differences, we can derive that to enhance users' interest in an SVE, providing mixed avatar appearances in an ambient crowd might be beneficial \citep{latoschikNotAloneHere2019}. In group situations, on the other hand, where the virtual others are in a realm that might enable direct interaction, even if it was framed as not allowed, a realistic appearance may increase the impression of interaction possibilities and, likewise, co-presence.

\subsubsection{Affective Appraisal}
\label{subsubsec:appraisal}
We  \replaced[id=R2]{exploratively evaluated the impact of virtual others' realism, and their congruence with the self-avatar's realism on affective appraisal and the Uncanny Valley effect (E1.2). Therefore, we investigated}{conducted an exploratory evaluation to investigate}
whether groups of virtual others, comprising varying numbers of realistic or stylized virtual humans, exhibited different levels of humanness and whether they appeared eerie to the users. Supporting the effectiveness of our manipulation, virtual others were perceived as more human when they included a higher proportion of realistic virtual humans in the group's configuration. Further, we did not observe any differences in the perceived eeriness attributed to the groups' configurations \added[id=R2]{for both self-avatar types}. We deduce that we did not encounter an uncanny valley effect for the virtual others and assume that the perceived eeriness did not interfere with our overall results. Simultaneously, there is no evidence to suggest that incongruent styles between the self-avatar and the virtual others led to an eerie perception of virtual others.\looseness=-1



\subsection{(In)Congruencies and the Perception of the Self-Representation}
\label{subsubsec:discussVBO}

\subsubsection{The Self-Avatar’s Congruence with the Participant}
\added[id=R2]{Concerning the self-avatar's congruence with the participant, we hypothesized the manipulation of the self-avatar’s congruence with the participant to lead to significantly higher scores in VBO and self-identification for realistic self-avatars (H2.1), while we expected no significant differences in agency, change and self-location (H2.2).}
In line with H2.1\deleted[id=R2]{and H2.3}, realistic self-avatars resulted in a significant increase in self-identification and VBO, measured both in VR and after all VR exposures (post-VR). We attribute these differences between the perception of realistic and stylized self-avatars to the influence of two higher-order cues contributing to the cognitive processing of the self-avatar's congruence with the participant. First, consistent with prior research suggesting that avatar realism enhances self-identification and VBO~\citep{latoschikEffectAvatarRealism2017, salageanMeetingYourVirtual2023}, a (more) realistic self-avatar may have provided visual cues that created the impression of sufficient human likeness congruent with real-world knowledge on human anatomy and composition \citep{lugrinAnthropomorphismIllusionVirtual2015}. 
Secondly, our results align with previous studies that have highlighted personalization as a key factor influencing VBO and self-identification~\citep{waltemateImpactAvatarPersonalization2018, salageanMeetingYourVirtual2023}. We assume that the scan-based personalization process employed for realistic self-avatars offered a higher degree of truthfulness compared to the customization process used for stylized self-avatars. Eventually, participants perceived a greater similarity between themselves and the realistic self-avatars and potentially attributed more personal characteristics to them~\citep{wolf2022observationdistance}. 
%
Interestingly, in contradiction to H2.2, realistic self-avatars also resulted in increased self-location, an effect we measured after the last VR exposure. We consider the named visual cues, to likely also have enhanced the spatial experience of being inside a virtual body. However, we did not observe the same effect in the repeated in-VR measurements, limiting the significance of the finding and leading us to propose further investigation.
Overall, we assume increased realism and truthfulness have resulted in visual cues congruent with the participants' real-world experiences concerning their own physical body, presuming one can identify with the virtual body, enhancing the sense of being inside and having it as one's own. \looseness=-1

Furthermore, in line with H2.2, the self-avatar's congruence with the participant did not yield differences in agency and change. Concerning agency, we posit that our technical system provided comparable embodiment configurations for both self-avatar types and, consequently, the same contingencies for controlling the virtual body during active movement~\citep{kilteniSenseEmbodimentVirtual2012}. We suggest the sense of agency over the virtual body primarily arose from the congruence of visuomotor and proprioceptive cues and, therefore, was less susceptible to manipulation from top-down influences. These results align with the recent meta-analysis by~\cite{mottelsonSystematicReviewMetaanalysis2023}, indicating limited effects of the avatar's appearance on agency. In terms of change, our findings indicated overall low values, which might be attributed to our efforts to minimize disparities between participants' actual body appearances and their self-avatars through the process of individualization, eventually leading to negligible alterations in the overall body schema.

\subsubsection{The Congruence Between the Self-Avatar and Virtual Others}
An exploratory evaluation of the congruence between self-avatars' realism and the virtual others' realism revealed notable effects on the perception of self-representation (E2.1). Interestingly, only for those participants embodying a stylized self-avatar, being together with an incongruent group of realistic virtual others led to lower ratings of self-location. A comparable pattern occurred for self-identification as both configurations of virtual others containing incongruent realistic virtual humans, i.e., mixed and realistic, seemed to hinder the process of identifying with the stylized self-avatar. Assuming self-location as well as self-identification to be concerned with the relationship between one's self and one's body~\citep{kilteniSenseEmbodimentVirtual2012}, it is noteworthy that higher-order (in)congruent visual cues that are not within the ego-central referential frame of one's (virtual) body, can have a\replaced[id=R1]{n}{ (negative)} effect on self-location and identification. The presence of realistic virtual others in the virtual environment may have accentuated the contrast between visual cues in rendering realistic virtual humans (others) and the stylized self-representation (self-avatar) not resembling realistic geometry and textures. This increased awareness may have further highlighted the incongruence between the participants' real-world bodies, as they are realistic per se, and the embodied stylized digital self-representation. The interaction plot (\autoref{fig:Results}) also indicates a similar tendency for VBO; however, the interaction was not statistically significant ($p = .084$). We are not aware of a comparable finding that the congruence between the realism of self-avatars and virtual others significantly impacts VBO, self-location, or identification with self-representation. It is critical to note that the questions used for self-location and self-identification~\citep{fiedler2023selfidentification} have not yet been validated. Therefore, our findings necessitate additional investigation to draw further conclusions. \looseness=-1

 In contrast to marginal indications from  \cite{latoschikEffectAvatarRealism2017}, we did not reveal an impact of virtual others' realism on participants' change in the self-perceived body schema. Besides differences in the executed tasks and the mirror exposure time between the named study and our experimental procedure, this discrepancy may be attributed to the individualized nature of our self-avatars, sharing fewer dissimilarities with participants' physical bodies, as opposed to the generalized avatars used in the work of \cite{latoschikEffectAvatarRealism2017}. This individualization may have resulted in fewer alterations to the overall body schema, potentially diminishing or eliminating the influence of others and their congruence with the self-avatar on the change measure.\looseness=-1

\subsection{(In)Congruencies and Spatial Presence}
In line with H3, the realism of the avatar did not lead to differences in spatial presence, nor did the style configuration of virtual others and their congruence. First, this finding aligns with the prevailing conceptualization of spatial presence as being primarily driven by bottom-up factors, influenced by the system's immersion \citep{slaterFrameworkImmersiveVirtual1997, slaterSeparateRealityUpdate2022}, a factor
we consistently maintained for all conditions. 
Secondly, in relation to the CaP model, our results demonstrate comparable rendering of congruent spatial cues across all avatars and configurations of virtual others. The manipulation of higher-order visual cues pertaining to the realism and individualization of virtual humans appeared to predominantly influence the plausibility of one's virtual body as well as the group of virtual others. 
However, these manipulations contributed less to a general sense of \enquote{being there} within the virtual environment. The recent experiment by \cite{salageanMeetingYourVirtual2023} supports our observations for the self-avatar, as the authors reported personalization, realism, or their interaction in self-avatars not to affect presence. 
\added[id=R1]{Interestingly, we provide evidence for a distinction between self-location, referring to the relationship between one’s
self and one’s body \citep{kilteniSenseEmbodimentVirtual2012} and spatial presences, concerned with the relationship
between one’s self and the environment \citep{wirthProcessModelFormation2007}. 
Our results indicate that manipulating the visual cues of the self-avatar and its congruence with virtual others impacted one’s spatial experience of being inside a virtual body rather than being inside a virtual environment.}
However, it's worth noting that there is conflicting evidence in the literature, with some studies suggesting that spatial presence may indeed benefit from realism in avatars \citep{weidnerSystematicReviewVisualization2023} as well as from personalization \citep{waltemateImpactAvatarPersonalization2018}. 
We recommend future work to delve deeper into exploring the relationship between the sense of being there in the virtual environment, the appearance of virtual humans, and the resulting congruence of spatial cues. This investigation might not only list related studies but also compare effect sizes and system configurations, which could serve as a valuable extension to the review of \cite{weidnerSystematicReviewVisualization2023}.\looseness=-1

\subsection{\added[id=R2]{The Impact of Realism in a Group of Virtual Humans}}
\added[id=R2]{
A substantial body of research has been concerned with understanding how the realism of avatars and agents impacts users' virtual experiences and their evaluation of co-located virtual others \mbox{\citep{weidnerSystematicReviewVisualization2023, nowakAvatarsComputermediatedCommunication2018}}. Yet, there has not been further investigation into the effects of being co-located with multiple virtual humans of different styles in an avatar-mediated VR setting, considering both the realism of co-located virtual humans, the realism of the self-avatar, and their (in)congruencies. Our results emphasize the benefits of a realistic appearance for a group of virtual humans. First, for virtual others, we identified that the group's overall realism, i.e., its congruence with participants' real-life experiences, benefits the VR experience in terms of co-presence and perceived interaction possibilities, while the internal congruence of the group's style configuration did not yield significant differences. Second, a realistic self-avatar congruent with the participants' real-world experiences concerning their own physical bodies was important for identifying with and accepting the virtual body as one's own. Lastly, only for stylized self-avatars did incongruencies between the self-avatar and the virtual others' configuration have an adverse effect on the relationship between one's self and body. This might even indicate that realistic self-avatars can help to adapt to incongruencies between the realism of self and others. We thus infer that the further development and increasing accessibility of technologies to provide lifelike avatars, e.g., through 3D reconstruction, can help to improve the VR experience in virtual group settings.}

\subsection{Limitations \added[id=R2]{and Future Work}}
\label{subsec:limitation}

In order to evaluate the implications of our results, we will outline certain limitations of our study and use them as a basis for future research. First, in an approach to ensure high experimental control and consistency of behavior across all configurations of virtual others, we decided to limit the interactivity between participants and the virtual others. Therefore, we implemented a rule-based, event-driven approach to generate virtual others' behavior~(see \autoref{subsec:virtualOthers}), not allowing for facial expressions and direct interaction. \replaced[id=R2]{While the work of \mbox{\citep{volonteEmpiricalEvaluationVirtual2018}} indicates comparable visual attention between conversational and non-conversational animations in virtual humans, our approach}{This} may have limited the ecological validity \citep{baumeisterEncyclopediaSocialPsychology2023} of the agents' displayed behavior, \added[id=A]{and may has conveyed false affordances, i.e., \enquote{possibilities for action and interactions that seem achievable but that are actually not carried out by the simulation} \citep[p. 3]{dufresneUnderstandingImpactCoherence2024}. This limitation was} also \deleted[id=A]{was} reflected in low values on the impression of interaction possibilities. We suggest future work to deepen insight by utilizing different styles of \deleted[id=R2]{human-controlled} virtual others \deleted[id=R2]{(avatars)} \replaced[id=R2]{in an extended scenario with social interaction, e.g. by modeling robust virtual human conversations \mbox{\citep{rossenHumanCenteredDistributedConversational2009}}, enabling AI-controlled virtual humans \mbox{\citep{wienrichEXtendedArtificialIntelligence2021}}, or implement}{in} a real social interaction between multiple users of an SVE.

Second, we provided different types of individualization for the types of virtual humans to shape the manipulation of the self-avatar's congruence with the participant's physical body. 
We assume that the personalization process resulted in a higher level of truthfulness compared to the customization process, as the personalization objectively resembled more details of the participant's actual appearance than customization. However, the two methods also differ in how participants have agency over their avatar's creation. While the personalization process primarily aims to mirror the current state of the participant's appearance, customization empowers participants to create their avatars based on their own goals, which may or may not prioritize a resemblance to reality. To maintain a reasonable level of consistency, we explicitly instructed the participants to choose configurations similar to their physical appearance. 
Future work should carefully consider the impact of differences in agency over the individualization process on an individual's VR experience. \looseness=-1



Lastly, given a sample size of N = 48, our analyses may not have had the intended power to reveal all hypothesized effects. A post hoc sensitivity analysis~($\alpha=0.05,\, 1-\beta = 0.8,\, N = 48$) using G*Power version 3.1.9.6~\citep{faulPowerFlexibleStatistical2007} revealed that a two-grouped (self-avatar) mixed ANOVA with three repeated measures (virtual others) would be sensitive to medium to large effects for between-group effects~($f = 0.34$) and small-to-medium effects for within-group and within-between-group interaction effects~($f = 0.19$)~\citep{cohenStatisticalPowerAnalysis2013}. 
Likewise, for the measures after all VR exposures, a post hoc sensitivity analysis for a t-test with independent samples ~($\alpha=0.05,\, 1-\beta = 0.8,\, N_1 = 24, N_2 =24$) would be sensitive to medium to large effects~($d = 0.73$). We, cannot rule out a type 2 error, especially for small to medium between-group effects. 

\subsection{Implications}
Our findings carry important implications for the design of social virtual environments (SVEs) incorporating virtual humans of varying degrees of realism and anthropomorphism. Overall, participants showed remarkable flexibility in accepting varying styles in a mixed group of virtual humans and its impact on the VR experience. \added[id=R2]{We assumed a homogeneous group configuration to be more congruent than a mixed one, an (in)congruence that can be accessed by directly comparing others within the VE. However, } these aspect of virtual others, namely their homogeneity of group configuration, did not yield differences in our qualia space, offering enormous versatility for the design of shared virtual applications. \added[id=R2]{We infer that mixing styles in a group of virtual others does not per se lead to a less plausible user experience.} However, another congruence in virtual others proved important; their lifelikeness or realism, respectively. 
Following our assumption, realistic virtual others may be congruent with the participant’s real-world experiences and expectations, elevating co-presence and the impression of interaction possibilities. \replaced[id=R2]{We propose considering realistic appearances for virtual others to enhance the realism and effectiveness of interactions \mbox{\citep{kyrlitsiasSocialInteractionAgents2022}}}{. 
This can be of high relevance}, e.g., in collaborative or educational workspaces, eventually fostering a more natural and productive immersive social interaction \citep{Orel2022, scavarelliVirtualRealityAugmented2021}. Furthermore, for self-representation, applications can benefit from congruence with the participant’s real-world experiences. Within the referential frame of one's physical body, a realistic and truthful self-avatar can enhance users' sense of owning and identification with the virtual body, which, e.g., might be particularly relevant in the therapeutical context supporting patients in developing realistic self-perception and a positive body image~\citep{dollingerViTraSVirtualReality2019}. However, relating to certain use cases, also avatars incongruent or \enquote{dissimilar} with the user's physical appearance could help to solve incongruencies within other referential frames, e.g., by establishing consistency between the self-avatar and other elements of the VE, or the overall framing or habituation of the experience \citep{cheymolMyRealBody2023a}. Lastly, the incongruences between the self-avatar of stylized appearance and realistic virtual others could lead to an altered self-perception with potential negative impacts on the relationship between our self and body. Considering this effect in the development of SVEs, future work might aim at locally transforming the style of all virtual others to a stylized appearance only for those participants embodying a stylized avatar. However, implementing such a transformation without the participant's knowledge can raise ethical concerns, and participants should be actively informed about the potential modifications to their virtual environment. \looseness=-1

\subsection{Conclusion} 
\label{subsec:conclusion}
Virtual humans play a pivotal role in a wide range of social virtual environments, and a remarkable body of research on the visualization of avatars and agents highlights the impact of virtual humans' visual cues contributing to a user's individual VR experience. Our work aligns with current theories and models, emphasizing the role of congruence and plausibility in influencing users' virtual experiences and effects, by focusing on the intricate dynamics emerging from (in)congruent styles of a group of virtual humans, including multiple co-located others~(agents), and one's digital self-representation~(avatar). We indicate groups of virtual others of higher realism to increase the feeling of co-presence and the impression of interaction possibilities, while (in)congruencies in the homogeneity of the group did not cause considerable effects. Furthermore, realistic self-avatars congruent with participants' own physical bodies yielded notable benefits for virtual body ownership and self-identification with the digital representation. Notably, the incongruence between a stylized self-avatar and a group of realistic virtual others resulted in diminished ratings of self-location and self-identification, suggesting an adverse effect on the relationship between one's self and body. In conclusion, a group of virtual humans varying in realism can result in a multitude of (in)congruent visual cues impacting a VR experience. Therefore, considering these (in)congruencies to tailor a virtual experience to the application's purpose and target audience constitutes an important yet challenging task to which we contribute empirical evidence and the discussion of their implications.  \looseness=-1

\section*{Conflict of Interest Statement}

The authors declare that the research was conducted in the absence of any commercial or financial relationships that could be construed as a potential conflict of interest.

\section*{Acknowledgments}
We thank Lena Holderrieth, Florian Kern, and Jonathan Tschanter, who are with the Human-Computer Interaction Group of the University of Würzburg, for their help in developing the Unity application and data collection. A preprint of this work has been published on March 11th, 2024 \citep{malAmOddOne2024}. 

\section*{Funding}
This research has been funded by the German Federal Ministry of Education and Research (BMBF) in the projects ViTraS (Grant 16SV8219 and 16SV8225) and ViLeArn More (Grant 16DHB2214), by the German Federal Ministry of Labour and Social Affairs (BMAS) in the project AIL AT WORK (Grant DKI.00.00030.21), and by the Bavarian State Ministry For Digital Affairs in the project XR Hub (Grant A5-3822-2-16). It was further supported by the Open Access Publication Fund of the University of Würzburg. Erik Wolf gratefully acknowledges a Meta Research Ph.D. Fellowship. \looseness=-1

\section*{Data Availability Statement}
The raw data supporting the conclusions of this article will be made available by the authors, without undue reservation.

\section*{Ethics Statement}
The studies involving humans were approved by the Ethics Committee of the Institute of Human-Computer-Media at the Faculty of Human Sciences of the Julius Maximilian University of Würzburg. The studies were conducted in accordance with the local legislation and institutional requirements. The participants provided their written informed consent to participate in this study. Written informed consent was obtained from the individual(s) for the publication of any potentially identifiable images or data included in this article.\looseness=-1

\section*{Author Contributions}
ML and DM conceptualized, discussed, and motivated the experimental design. DM took the lead in data collection, writing, reviewing, and editing the manuscript. ND supported in data analysis. EW and DM developed the Unity application. MB and SW provided the creation of personalized, realistic virtual humans. ML and CW initiated the idea and goals of the study, acquired financial support, and supervised and managed the project. All authors continuously provided constructive feedback and helped to write, edit, and refine the manuscript.

\bibliographystyle{template/frontiersinSCNS_ENG_HUMS} 
\bibliography{2024-vhp-group}

\end{document}